\PassOptionsToPackage{dvipsnames}{xcolor}
\documentclass[longbibliography,nofootinbib,aps,prx,superscriptaddress,onecolumn]{revtex4-2}

\usepackage{amsmath,amsfonts,amssymb,amsthm,dcolumn,bm,qcircuit,appendix,mathtools,thmtools,thm-restate,mathrsfs,pgfplots,soul,lipsum,graphicx,braket,enumitem,caption,subcaption,ragged2e}
\usetikzlibrary{calc}

\usepackage{tikz,tkz-euclide,tikz-3dplot}
\usetikzlibrary{positioning}

\usetikzlibrary{matrix,fit,backgrounds,3d,arrows, decorations.pathreplacing,shapes.geometric,3d, calc}

\def\BibTeX{{\rm B\kern-.05em{\sc i\kern-.025em b}\kern-.08em
    T\kern-.1667em\lower.7ex\hbox{E}\kern-.125emX}}
    
\usepackage[ruled,vlined]{algorithm2e}
\usepackage{algorithmic}

\newtheorem{theorem}{Theorem}
\newtheorem{lemma}{Lemma}

\newtheorem{remark}{Remark}
\newtheorem{definition}{Definition}
\newtheorem{method}{Algorithm}

\DeclareMathOperator{\Tr}{Tr}
\newcommand{\xbf}{\textbf{x}}

\newcommand{\Ibb}{\mathbb{I}}

\newcommand{\Rbb}{\mathbb{R}}

\newcommand{\Zbb}{\mathbb{Z}}
\newcommand{\Scal}{\mathcal{S}}
\newcommand{\norm}[1]{\left\lVert#1\right\rVert}
\newcommand{\abs}[1]{\lvert#1\rvert}

\usepackage{xcolor}
\usepackage{hyperref}

\hypersetup{
	colorlinks = true,
	linkcolor = [rgb]{0.70,0.13,0.13},
	citecolor = [rgb]{0.13,0.55,0.13},
	urlcolor = [rgb]{0.25, 0.41, 0.88}}
\usepackage[capitalize,nameinlink]{cleveref}

\begin{document}
\title{Towards quantum topological data analysis: torsion detection}

\author{Nhat A. Nghiem}
\email{{nhatanh.nghiemvu@stonybrook.edu}}
\affiliation{Department of Physics and Astronomy, State University of New York at Stony Brook, Stony Brook, NY 11794-3800, USA}
\affiliation{C. N. Yang Institute for Theoretical Physics, State University of New York at Stony Brook, Stony Brook, NY 11794-3840, USA}

\begin{abstract}
Topological data analysis (TDA) has become an attractive area for the application of quantum computing. Recent advances have uncovered many interesting connections between the two fields. On one hand, complexity-theoretic results show that estimating Betti numbers—a central task in TDA— is $\rm NP$-hard, indicating that a generic quantum speedup is unlikely. On the other hand, several recent studies have explored structured, non-generic settings and demonstrated that quantum algorithms can still achieve significant speedups under certain conditions. To date, most of these efforts have focused on Betti numbers, which are topological invariants capturing the intrinsic connectivity and “holes” in a dataset. However, there is another important feature of topological spaces: torsion. Torsion represents a distinct component of homology that can reveal richer structural information. In this work, we introduce a quantum algorithm for torsion detection—determining whether a given simplicial complex contains torsion. Our algorithm, assisted by a classical procedure (for diagonalizing small integer matrix) of low complexity, can succeed with high probability and offer polynomial speed-up (in non-oracle setting) to exponential speedup (in oracle-setting) over the best-known classical approach.

%Topological data analysis (TDA) has emerged as a promising candidate for the application of quantum computers. Recent progress has revealed many interesting aspects of TDA, particularly its interplay with quantum computing. On the one hand, there has been a complexity-theoretic result showing that estimating Betti numbers, a central problem in TDA, is $\rm NP$-hard. This result has imposed a strong barrier, suggesting that quantum speedup in a generic setting is not possible. On the other hand, a few recent efforts have undergone in-depth investigation into non-generic setting, and revealed that quantum speedup can still be possible in a structured scenario. These previous attempts have mainly focused on estimating Betti numbers, and such numbers are topological invariants that capture the intrinsic topological structure of the given dataset. However, there is another component, called \textit{torsion}, may associate with the given complex of interest. In this work, we propose a quantum algorithm for detecting torsion, i.e., determining whether the given complex has torsion. In certain setting, our quantum algorithm, assisted by a low-cost classical procedure, can succeed with high probability and yields exponential speedup compared to the classical counterpart. 

\end{abstract}

\maketitle

% \tableofcontents
% \newpage
% === INTRO === %
\section{Introduction}
As an application of quantum theory—a scientific revolution of the early twentieth century—quantum computation has emerged as a powerful computing paradigm. Considerable effort has been devoted to exploring its potential, leading to many landmark algorithms and results \cite{shor1999polynomial, grover1996fast, deutsch1985quantum, deutsch1992rapid, harrow2009quantum, aharonov2003adiabatic, lloyd1996universal, lloyd2013quantum, lloyd2020quantum, berry2007efficient, berry2012black, berry2014high, berry2015hamiltonian, berry2015simulating, berry2017quantum, childs2017quantum, mitarai2018quantum}. More recently, there has been a surge of interest in applying quantum computation to topological data analysis (TDA)—a relatively young but rapidly growing field that uses tools from algebraic topology to study large-scale, complex data. Although TDA offers powerful insights, its methods can be computationally expensive, motivating the search for faster alternatives. Quantum computing, with its ability to manipulate high-dimensional states efficiently, has been viewed as a promising route to overcoming some of these computational challenges.

The interplay between quantum computing and TDA has already produced several interesting results. Lloyd, Garnerone, and Zanardi \cite{lloyd2016quantum} introduced the first quantum algorithm (often called the LGZ algorithm) for estimating Betti numbers, one of the central invariants in TDA. Subsequent works \cite{berry2024analyzing, ubaru2021quantum, mcardle2022streamlined} improved the LGZ framework in various ways. Hayakawa et al. \cite{hayakawa2022quantum} proposed a quantum method for estimating persistent Betti numbers, a more refined invariant. Others have examined the complexity-theoretic aspects of Betti number computation \cite{schmidhuber2022complexity, crichigno2024clique}. Recent studies \cite{lee2025new, nghiem2023quantum} further demonstrate that, for certain classes of simplicial complexes, quantum algorithms can outperform previous approaches \cite{lloyd2016quantum, berry2024analyzing, ubaru2021quantum}.

While most of these works focus on Betti numbers, they capture only part of the topological information. For a given simplicial complex, the Betti numbers measure the rank of the free part of the homology group. However, homology also contains another component: torsion. Intuitively, Betti numbers count the number of independent “holes” or cycles that cannot be filled, whereas torsion measures those cycles that become trivial (i.e., contractible) after a finite number of traversals. Together, the free part and the torsion part give a more complete characterization of a topological space. For example, a sphere and a Klein bottle differ not only in their Betti numbers but also in that the Klein bottle has torsion while the sphere does not. Yet, most existing TDA algorithms, classical or quantum, typically work over $\Rbb$ or $\Zbb_2$, effectively hiding torsion information. Readers interested in a more formal discussion of this issue are referred to Appendix~\ref{sec: roleofcoefficients}, which explains how the choice of coefficients can obscure torsion.

Motivated by this gap, our work explores the potential of quantum computation for torsion detection. Specifically, given a simplicial complex $K$, we ask: Does $K$ contain torsion? Our algorithm draws on several mathematical tools. The key theoretical foundations are classical results from algebra, including the structure theorem for finitely generated abelian groups and the universal coefficient theorem, both of which imply that torsion is reflected in the dimensions of certain homology groups. Thus, our strategy reduces to estimating the dimensions of these groups—a task for which we propose a quantum procedure. Our method builds on recent advances in state preparation, block-encoding, and the quantum singular value transformation (QSVT) framework. We show that, with suitable input, a quantum computer can detect torsion significantly faster than known classical algorithms.

The rest of the paper is organized as follows. Section~\ref{sec: topologicaldataanalysis} provides a brief introduction to TDA. Section~\ref{sec: surveyqtda} surveys the state-of-the-art quantum algorithms for TDA. Section~\ref{sec: ourcontribution} presents our main contribution: a quantum algorithm for torsion detection. We conclude in Section~\ref{sec: conclusion} with broader reflections and future directions. Several appendices contain detailed material: Appendix~\ref{sec: summaryofnecessarytechniques} reviews block-encoding and related techniques; Appendix~\ref{sec: reviewalgebra} summarizes algebraic background; Appendix~\ref{sec: reviewofalgebraictopology} reviews topological concepts; Appendix~\ref{sec: roleofcoefficients} elaborates on the role of coefficient fields; Appendix~\ref{sec: keyinsight} develops the main insight behind our method; Appendix~\ref{sec: quantumalgorithm} gives the full algorithm; and Appendix~\ref{sec: complexityanalysis} provides a detailed complexity analysis.

\section{Topological Data Analysis}
\label{sec: topologicaldataanalysis}
In this section, we provide an overview of topological data analysis (TDA), particularly its mathematical background. We keep things concise, intuitive and neat here and refer the interested readers to the Appendix \ref{sec: reviewalgebra} and \ref{sec: reviewofalgebraictopology}  for a more formal definitions.

Topological data analysis is a data analysis tool based on algebraic topology. Topology studies shape (e.g., how many holes a configuration have) while algebra studies arithmetic (what are properties of a symbolic set under certain operational rules). Algebraic topology is a mathematical framework that studies the topological space using algebraic methods. This is partly possible because, from the view of topology, certain objects are ``the same'', or more precisely, equivalent, up to a special type of transformation called homeomorphism. Intuitively, it is a deformation, or stretching without breaking, e.g., see the figure below for a simple illustration in 1-dimension. 
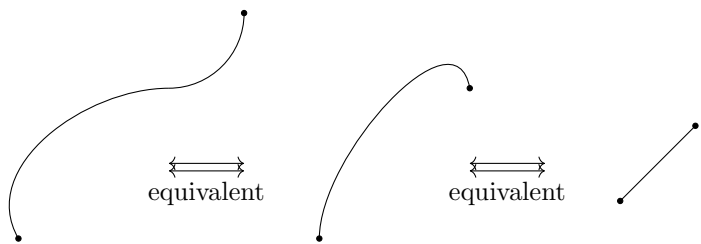
\begin{center}
\begin{tikzpicture}
    \node (A) at (1,1) {};
  \node (B) at (-1,-1) {};
  \filldraw (A) circle (1pt);
  \filldraw (B) circle (1pt);
  \draw (-1,-1) to[out = 90, in = 100] (1,1);
  \filldraw (3,-0.5) circle (1pt);
  \filldraw (4,0.5) circle (1pt);
  \draw (3,-0.5) -- (4,0.5);
  \draw[<->]  (1.0,0) -- (2.0,0) ;
  \draw[<->] (1.0, -0.1) -- (2.0, -0.1);
  \node at (1.5, -0.4) {equivalent};

    %\filldraw (-3,1) circle (1pt);
    \filldraw (-5, -1) circle (1pt);
    \filldraw (-2, 2) circle (1pt);
    \draw (-5, -1) to[out = 120, in = 180] (-3,1);
    \draw (-3,1) to[out = 0, in = 270] (-2,2);
     \draw[<->]  (-3,0) -- (-2.0,0) ;
  \draw[<->] (-3.0, -0.1) -- (-2.0, -0.1);
   \node at (-2.5, -0.4) {equivalent};
\end{tikzpicture}    
\end{center}
The above example features the fact that a curve, no matter how its shape look like, is equivalent to a straight line. The higher dimensional analog of this example can be carried out in a straightforward way. In 2-dimensional case, for instance, it can be seen that any surface domain is equivalent to a triangle. Thus, by reducing the seemingly complicated configuration to simpler ones, one can hope to gain its topological information easier.

\begin{figure}[]
\centering
\begin{tikzpicture}[scale = 1.7, every node/.style={font=\small}]
    % 0-simplex
    \filldraw (-5,-1) circle (1pt);
    \node[above left] at  (-5,-1) {$v_0$}; 
    \node[below right] at  (-5.5,-1) {$0$-simplex}; 
    
    % 1-simplex
    \draw (-3.5,-1) -- (-2.5,-1);
    \filldraw (-3.5,-1) circle (1pt);
    \filldraw (-2.5,-1) circle (1pt);
    \node[above] at (-3.5,-1) {$v_0$};
    \node[above] at (-2.5,-1) {$v_1$}; 
    \node[below] at (-3.0,-1) {$1$-simplex}; 
    
    % 2-simplex
    \coordinate (v0) at (-1,-0.5); 
    \coordinate (v1) at (-1.6, -1.5); 
    \coordinate (v2) at (-0.4, -1.5); 
    \draw[fill=blue!10] (v0)--(v1)--(v2)--cycle; 
    \filldraw (v0) circle (1pt);
    \filldraw (v1) circle (1pt);
    \filldraw (v2) circle (1pt);
    \node[above] at (v0) {$v_0$};
    \node[below left] at (v1) {$v_1$};
    \node[below right] at (v2) {$v_2$};
    \node[below] at (-1.0, -1.8) {$2$-simplex}; 
    
    % 3-simplex (tetrahedron)
    \coordinate (P0) at (1.5,-0.4);
    \coordinate (P1) at (0.7,-1.4);
    \coordinate (P2) at (1.9,-1.6);
    \coordinate (P3) at (1.9,-1.0);
    \draw[fill=blue!10] (P0) -- (P1) -- (P2) -- cycle; % base face
    \draw[thick] (P0) -- (P1) -- (P2) -- (P0);
    \draw[fill=blue!10,thick] (P0) -- (P3) -- (P2);
    \draw[dashed] (P1) -- (P3);
    \filldraw (P0) circle (1pt);
    \filldraw (P1) circle (1pt);
    \filldraw (P2) circle (1pt);
    \filldraw (P3) circle (1pt);
    \node[above] at (P0) {$v_0$};
    \node[below left] at (P1) {$v_1$};
    \node[below right] at (P2) {$v_2$};
    \node[right] at (P3) {$v_3$}; 
    \node[below] at (1.3, -1.8) {$3$-simplex}; 
\end{tikzpicture}
\caption{\justifying \textbf{Illustration of standard simplexes.} From left to right: a point ($0$-simplex), a line segment ($1$-simplex), a filled triangle ($2$-simplex), a filled tetrahedron ($3$-simplex). Each $r$-simplex is formed by $(r{+}1)$-geometrically independent vertices in Euclidean space.}
\label{fig: simplex}
\end{figure}
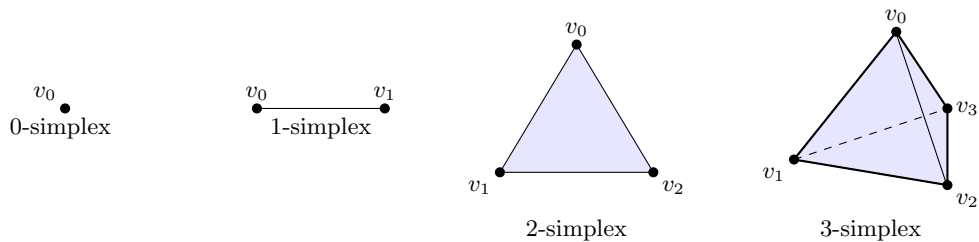

These objects, including line, triangle, and higher-dimensional one, are the building blocks of algebraic topology, which are named simplexes. Figure \ref{fig: simplex} illustrates 0-simplex, 1-simplex, 2-simplex and 3-simplex. Higher-dimensional simplexes are built in a similar way: an $r$-simplex contains $(r+1)$ geometrically independent points. A simplicial complex is built by gluing simplexes so that the intersection between arbitrary two simplexes is either another simplex or empty. 
Earlier, we have mentioned the notion of homeomorphism. If there is a homeomorphism between a topological space $\mathcal{X}$ and a simplicial complex $K$, then $K$ is called a triangulation of $\mathcal{X}$. The key philosophy of algebraic topology is that, by associating/approximating a continuous object (e.g., topological space) with discrete objects (simplicial complex -- essentially a collection of simplexes), there exists a correspondence between their topological feature of the former and algebraic feature of the latter. By analyzing the algebraic structure of the discrete representation, the topological feature of the original space can be inferred.

To analyze such an algebraic structure, it is necessary to assign certain algebraic properties to the given complex. Throughout our work, we use $\Scal_r$ to denote the set of $r$-simplexes in the complex $K$ of interest. Then a (formally) linear combination of these $r$-simplexes form a $r$-chain. The collection of these chains thus form a $r$-chain group/space $C_r$. In particular, there is an important operation called \textit{boundary operator}. Roughly speaking, it decomposes a simplex into a (formal) sum of its boundary. For instance, the boundary operator acts on 2-simplex as follows.
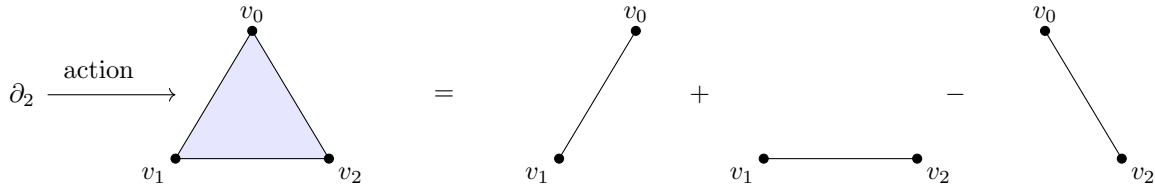
\begin{center}
    \begin{tikzpicture}[scale = 1.7]
    % 2-simplex
    \coordinate (v0) at (-1,-0.5); 
    \coordinate (v1) at (-1.6, -1.5); 
    \coordinate (v2) at (-0.4, -1.5); 
    \draw[fill=blue!10] (v0)--(v1)--(v2)--cycle; 
    \filldraw (v0) circle (1pt);
    \filldraw (v1) circle (1pt);
    \filldraw (v2) circle (1pt);
    \node[above] at (v0) {$v_0$};
    \node[below left] at (v1) {$v_1$};
    \node[below right] at (v2) {$v_2$};
    %\node[below] at (-1.0, -1.8) {$2$-simplex}; 
    \node at (-2.8, -1) {$\partial_2 $};
    \draw[->] (-2.6,-1) -- (-1.6,-1);
    \node at (-2.2, -0.8)  {action}; 
    \node at (0.5, -1) {=};
    \filldraw (2, -0.5) circle (1pt); 
    \node[above] at (2,-0.5) {$v_0$};
    \filldraw (1.4, -1.5) circle( 1pt);
    \node[below left] at (1.4,-1.5) {$v_1$};
    \draw (2,-0.5) -- (1.4,-1.5);
    \filldraw (3, -1.5) circle (1pt);
    \node[below left] at (3,-1.5) {$v_1$};
    \filldraw (4.2, -1.5) circle (1pt);
    \node[below right] at (4.2,-1.5) {$v_2$};
    \draw (3,-1.5) -- (4.2, -1.5); 
    \filldraw (5.2, -0.5) circle (1pt);
    \node[above] at (5.2,-0.5) {$v_0$};
    \filldraw (5.8, -1.5) circle (1pt);
    \node[below right] at (5.8,-1.5) {$v_2$};
    \draw (5.2,-0.5) -- (5.8, -1.5);
    \node at (2.5, -1) { $+$};
    \node at (4.5, -1) {$-$};
\end{tikzpicture} 
\end{center}
where the plus (+) or minus (-) sign depends on the order of the vertex removed. The generalization to higher-dimensional simplexes is straightforward, obeying the following formula: 
\begin{equation}
\partial_r [p_0, p_1, \ldots, p_r] = \sum_{i=0}^r (-1)^i [p_0, p_1, \ldots, \hat{p_i}, \ldots, p_r],
\end{equation}
The action of the boundary operator $\partial_r$ in the $r$-chain (linear combination of $r$-simplexes) is enacted by linearity. While a more detailed treatment of algebraic topology can be found in \cite{hatcher2005algebraic} (see also Appendix \ref{sec: reviewofalgebraictopology}), we remark that, as a fundamental result in algebraic topology, the topological structure of the given complex is encoded in the algebraic structure of the boundary operators. Specifically, suppose that we work over real field $\Rbb$, then $\partial_r$ becomes a linear map, and we have that the $r$-th homology space is defined as 
\begin{align}
    H_r \equiv \rm Ker (\partial_r) / \rm Im (\partial_{r+1})
\end{align}
The so-called $r$-th Betti number $\beta_r$ is defined as:
\begin{align}
    \beta_r = \dim H_r = \dim \rm Ker\left(  \Delta_r \right) 
\end{align}
The operator $\Delta_r=  \partial_{r+1} \partial_{r+1}^\dagger + \partial_r^\dagger \partial_r $ is named combinatorial Laplacian operator. Intuitively, $\beta_r$ captures the number of $r$-dimensional holes in the complex. Thus, as we mentioned earlier, by algebraically analyzing the boundary operators, we can deduce the topological space, e.g., revealing how many holes it has.

To see how algebraic topology can be relevant to data analysis, we point out that, in reality, data usually come as vectors, or points in the Euclidean space, with mutual connectivity provided, e.g., a graph network, triangulated mesh, etc. Such a network can be treated as a simplicial complex, and thus, algebraic topology can be applied. In reality, there can be many data points, and they can be very high-dimensional. Thus, it presents a nontrivial challenge for analysis. For example, provided a dataset, a question one may ask is: how does the dataset look like? We have pointed out from the beginning that topology cares about shape, and thus, TDA emerges as a suitable tool for answering this question. Indeed, there have been many efforts to apply TDA \cite{carlsson2021topological, bukkuri2021applications, singh2023topological, patania2017topological,seversky2016time,bubenik2015statistical}. These applications span many areas, from biomedicine, computer vision, to financial analysis. Their successes have affirmed that TDA is a powerful data analytical tool, and that it can complement nicely to the well-known ones, such as machine learning, statistical methods, etc.

\section{A Survey of Quantum Topological Data Analysis}
\label{sec: surveyqtda}
Quantum algorithms for topological data analysis (briefly called quantum TDA), specifically estimating Betti numbers, began with the work in \cite{lloyd2016quantum}, a.k.a LGZ algorithm. Their algorithm contains the following key components as well as related assumptions.
\begin{enumerate}
    \item \textbf{Encoding simplexes into computational basis states.} Let the simplicial complex of interest $K$ having $N$ data points, denoted as $v_0,v_1,v_2,...,v_N$. Then a $r$-simplex $\sigma_r$ $\in K$ is encoded into a $N$-qubit string of Hamming weight $r+1$ $\ket{\sigma_r}$ as follows: if $v_i \in \sigma_r$ then the value of $i$-th bit is 1; otherwise, it is 0. 
    \item \textbf{Oracle for verifying simplexes.} For any $r$ ($1 \leq r \leq N$), there is an oracle $O_r$ that acts as follows:
    \begin{align}
    O_r \ket{0} \ket{\sigma_r} = \begin{cases}
        \ket{1} \ket{\sigma_r} \text{\ if $s_r \in $ K  }\\
        \ket{0} \ket{\sigma_r} \text{ otherwise} 
    \end{cases}
\end{align}
    \item \textbf{Entry-computability of boundary operators $\{ \partial_r\}_{r=1}^N$. } There are specifically two oracles. The first one computes the location of nonzero entries (for a given row/column). The second one computes the value of those entries. This assumption is essentially analogous to those appeared the context of Hamiltonian simulation \cite{berry2007efficient,berry2012black,berry2014high,berry2015hamiltonian}. In fact, the LGZ algorithm uses these quantum simulation algorithms as a subroutine. 
\end{enumerate}

With the above inputs, LGZ algorithm combines several well-known ones, including Grover's search algorithm (multi-solution version), quantum simulation, and quantum phase estimation, to construct a quantum procedure that estimates the $r$-th normalized Betti number $\frac{\beta_r}{|\Scal_r|}$ (where we remind that $\Scal_r$ is the set of $r$-simplexes in $K$), up to some additive accuracy. 

The LGZ algorithm has ignited many subsequent attempts, toward many directions. The work in \cite{ubaru2021quantum} proposed to replace the Grover's search subroutine within LGZ algorithm by a method called rejection sampling. In addition, they also proposed replacing the quantum phase estimation subroutine with the stochastic rank estimation method, which was first introduced in \cite{ubaru2016fast,ubaru2017fast}. These two replacements resulted in a certain improvement over the LGZ algorithm, and as claimed in \cite{ubaru2021quantum}, it is very suitable for near-term era. Recent work \cite{berry2024analyzing} provides an in-depth analysis of quantum advantage in estimating Betti numbers. In particular, they introduced new ways to replace Grover's search algorithm (with Dicke state preparation) and amplitude estimation (using Kaiser window). They also provided specific type of graphs/complexes for which exponential quantum speedup is possible. 

In addition to the two above, the work in \cite{hayakawa2022quantum, mcardle2022streamlined} also features another progress. Instead of estimating Betti numbers (as in \cite{lloyd2016quantum,berry2024analyzing, ubaru2021quantum}), the authors in \cite{hayakawa2022quantum, mcardle2022streamlined} focus on persistent Betti numbers and develop the corresponding quantum algorithms. We remind the reader that the Betti numbers capture the number of holes in a given complex. Persistent Betti numbers, on the other hand, capture the number of holes that survive over different \textit{filtration}. To get more insight about filtration, one can imagine that a simplicial complex is built on simplexes, which are again built on data points and pairwise connectivity. Filtration roughly means denser and denser connectivity. 

Aside from the algorithmic aspect, quantum TDA is also attracting attention from the complexity aspect. In particular, the work in \cite{schmidhuber2022complexity} has shown that, provided pairwise connectivity, computing Betti numbers is $\# \rm P$-hard, while estimating them is $\rm NP$-hard. A similar result obtained in \cite{crichigno2024clique}, showing that estimating Betti numbers $\rm QMA$-hard, while exactly computing them is $\rm BQP$-hard. These results have implied a strong barrier on the potential of quantum exponential advantage in TDA. However, the work \cite{gyurik2022towards} has shown that the problem of estimating normalized Betti numbers is classically hard, which leaves a certain hope for quantum TDA. In fact, recent efforts  \cite{lee2025new, nghiem2023quantum, nghiem2025hybrid} have shown that in structured settings, e.g., triangulated manifold \cite{nghiem2023quantum}, mutual relation between a pair of simplexes \cite{lee2025new}, or a graph with bounded degree \cite{nghiem2025hybrid}, quantum computers can still offer a significant advantage.

\section{Our Contribution}
\label{sec: ourcontribution}
The aforementioned works have featured a significant amount of interest in quantum TDA. Here, we make a relevant, yet fundamentally different progress toward this exciting direction. Instead of seeking the Betti numbers of a given complex, we are curious about the \textit{torsion} of such the complex. To elaborate on this concept, we recall that earlier we mentioned that a $r$-chain is formed by taking a (formal) linear combination of $r$-simplexes. The collection of these chains form the $r$-chain group/space $C_r$. If the coefficients of the linear combination are drawn from real field $\Rbb$, then $C_r$ behaves as a linear vector space, for which the Betti numbers $\beta_r$ can be estimated by probing the spectrum of corresponding Laplacian operators. In particular, we remark that all the works mentioned in the previous section are based on homology \textit{over $\Rbb$} (see also Appendix \ref{sec: reviewofalgebraictopology}). If instead of $\Rbb$, the coefficients are drawn from the integer $\Zbb$, which is a ring, then it leads to a distinct separation. 

First, over $\Zbb$, $C_r$ is no longer a vector space. Rather, it is an Abelian group (or $\Zbb$-module). Second, because of this, $\partial_r$ is no longer a linear mapping between vector spaces. Third, as a result, $H_r$ is no longer a space but an Abelian group. Since it is an Abelian group, there exists a fundamental result in algebra (see Theorem \ref{thm: structuredtheorem} in the Appendix \ref{sec: reviewalgebra}), showing that $H_r$ can be decomposed as:
\begin{align}
    H_r \cong \Zbb^{\beta_r} \oplus \left( \bigoplus_i Z_{r_i} \right)
\end{align}
where $r_i \mid r_{i+1}$ refers to the divisbility relation. We refer the readers to Appendix \ref{sec: reviewalgebra} for more a formal review of abstract algebra. The first part $\Zbb^{\beta_r} $ is called \textit{free part}, while the second part $ \left( \bigoplus_i Z_{r_i} \right)$ is \textit{torsion part}. To this end, we have sufficient information to state the key objective of our work, which is:
\begin{center}
    \textbf{Given a simplicial complex $K$, does it have torsion ?}
\end{center}
We remark a few things. In the setting $\Zbb$, $H_r$ is no longer a vector space and, therefore, the notion of $\rm dim \ H_r$ is no longer relevant. Rather, it is $\rm rank \ H_r$ that is of interest. However, quantum computation is built on quantum mechanics, which is naturally built on the complex field $\mathbb{C}$. Thus, it is of fundamental constraint that prevents existing quantum algorithms from handling the torsion part appearing in the above formula. 

Our solution to the above question and the corresponding quantum algorithm rely on a few insights. In the following, we attempt to summarize the main recipes and key steps. The detailed treatment shall be provided in the Appendix \ref{sec: roleofcoefficients}, Appendix \ref{sec: keyinsight} and Appendix \ref{sec: quantumalgorithm}. 

\subsection{Mathematical insight}
Let $H_r(Q)$ to denote the $r$-th homology group $H_r$ defined over $Q$, which can be a field (like real field $\Rbb$, complex field $\mathbb{C}$) or a ring (like $\Zbb$). The first recipe we use is the well-known result within algebra, called \textit{universal coefficient theorem}, which shows how the change in the choice of coefficients induces the change in the algebraic structure of the homology group:
\begin{align}
    H_r(Q) \cong H_r(\Zbb) \otimes_{\Zbb} Q  \oplus \rm Tor \left( H_{r-1} (\Zbb), Q  \right)
\end{align}
where the last term $\rm Tor \left( H_{r-1} (\Zbb), \Rbb   \right) $ is the \textit{torsion product functor} (see further Appendix \ref{sec: roleofcoefficients} for more details). If we choose $Q = \mathbb{F}_p \equiv \Zbb/p\Zbb$ for prime $p$, then we can turn $\mathbb{F}_p$ into a field, i.e., it is a finite field. As it is a field, the chain group $C_r$ becomes a vector space, and thus the homology group $H_r(\mathbb{F}_p)$ becomes a vector space. In the Appendix \ref{sec: roleofcoefficients} and \ref{sec: keyinsight}, using the above universal coefficient theorem and related properties of torsion product functor, we will prove the following:
\begin{align}
\rm dim \left( H_r(\mathbb{F}_p) \right) = \beta_r + t_r + t_{r-1} 
\end{align}
where $t_r, t_{r-1}$ is the total amount of the cyclic summands (of $k$-th homology group/space and $(k-1)$-th homology group/space, respectively) of order divisible by a prime $p$. At the same time, if we choose $Q= \Rbb$ as usual, then $H_r(\Rbb) \cong \Zbb^{\beta_r}$ and thus $\dim H_r(\Rbb) = \beta_r$, which is exactly the $r$-th Betti number. Thus, our strategy is, we estimate the dimension of the homology group over $\Rbb$, and $\mathbb{F}_p$ for various values of prime $p$. If they are the same, then it implies that there is no torsion at the order $r$ and $r-1$. Otherwise, there is. We remark a property that over finite field $\mathbb{F}_p$, it still holds that the dimension of homology group is equal to the dimension of the kernel of combinatorial Laplacian. However, we need to take into account the finite field, which results in that all arithmetic operations need to be done modulo $p$, e.g.,
\begin{align}
   \dim H_r(\mathbb{F}_p)  &=\dim \rm Ker \left(\Delta_r\right)^{\rm mod \ p}
\end{align}

\subsection{Necessary Recipes}

Before we outline our algorithm, we remark that a majority of our work relies on block-encoding. We refer the unfamiliar readers to the Appendix \ref{sec: summaryofnecessarytechniques} for a summary of formal definition as well as related operations involving block-encoding. 

We first describe the input model on which we shall work. Some existing works, such as \cite{lloyd2016quantum, berry2024analyzing, ubaru2021quantum, schmidhuber2022complexity, hayakawa2022quantum} assume an oracle that can verify the existence/non-existence of simplexes within the complex of interest. Particularly, the work \cite{berry2024analyzing} provides an explicit construction of such oracle, given the graph description of the interested complex. We name this setup oracle model. At the same time, recent results \cite{nghiem2023quantum, lee2025new} do not assume the oracle, and thus we name it non-oracle model. Rather, the input assumption  is the classical knowledge/description of a matrix that encodes the mutual relation among simplexes. More specifically, let $\Scal_r = \{ \sigma_{r_1}, \sigma_{r_2}, ... \}$ denotes the set of $r$-simplexes in the complex of interest $K$. For some $r$, let $\mathscr{D}_r$ denotes the matrix of size $ |\Scal_{r-1}| \times |\Scal_r|$, which encodes the structure of the given complex as follows. For a given column index, say $j$ (with $1 \leq j \leq |\Scal_r|$), we have:
\begin{align}
    (\mathscr{D}_r)_{ij} = \begin{cases}
        1 \text{ \ if $\sigma_{r_i} \cap \sigma_{(r-1)_j} \neq \emptyset$  } \\
        0 \text{ \ if $\sigma_{r_i} \cap \sigma_{(r-1)_j} = \emptyset$  } 
    \end{cases}
\end{align}
The assumption is that for any $r$, the entries of $\mathscr{D}_r$ are classically known/provided. As discussed in \cite{lee2025new}, a way to achieve this assumption is by direct specification, which is also partially suggested in \cite{schmidhuber2022complexity}. In the Appendix \ref{sec: quantumalgorithm}, we will describe this input model in greater detail, including an example for illustration. In the following, we proceed to describe our algorithm with the non-oracle model as input. Subsequently, we will show how the oracle model can also be easily integrated into our procedure. 

To proceed, we point out the following result regarding state preparation with elements drawn from a finite field, which is an important recipe for our subsequent construction: 
\begin{lemma}[Finite-field state preparation \cite{marin2023quantum, mcardle2022quantum, nakaji2022approximate})-- Appendix \ref{sec: proofofstatepreparation}]
\label{lemma: finitefieldstatepreparation}
    Given a $N$-dimensional quantum state $\ket{\Phi}= \frac{1}{||\xbf||} \sum_{i=1}^N x_i \ket{i}$  where $||\xbf|| = \sqrt{\sum_{i=1}^N x_i ^2 }$, and each $x_i$ is drawn from a finite field, e.g., $\mathbb{F}_p$. Then there is a quantum procedure that prepares $\ket{\Phi}$, with circuit complexity $\mathcal{O}\left( \log N \right) $ and extra $\mathcal{O}(1)$ ancilla qubits. 
\end{lemma}
Generally speaking, the above quantum procedure directly applies the state preparation protocols in \cite{mcardle2022quantum, marin2023quantum, nakaji2022approximate}. For instance, according to Theorem 1 in \cite{mcardle2022quantum}, there is an efficient procedure to prepare the state of the form $ \ket{\Phi_f} \sim \sum_{ i=-\frac{N}{2}}^{\frac{N}{2 } -1} f \left( \frac{2a i}{N} \right)  \ket{i}$ (where $\sim $ hides the normalization factor), where $f: [-a,a] \longrightarrow \Rbb$ being some preferably smooth function (albeit a non-smooth function also works, as discussed in \cite{mcardle2022quantum}). By choosing any $f$ that satisfies $f(\frac{2a i}{N} ) = x_i$, then the desired state $\ket{\Phi}$ can be prepared. The circuit complexity of this method is $\mathcal{O}\left( \log N\right) $, and at most 3 ancilla qubits. As a comment, in principle, any function $f$ with values as desired can be used. In Appendix \ref{sec: proofofstatepreparation}, we will describe an effective approach to obtain the desired $f$ in practice. The general idea is to promote these amplitudes $\{x_i\}_{i=1}^N$ to the corresponding data points $\{ (2a i/N,x_i) \}_{i=1}^N \in \Rbb^2$. By connecting two consecutive data points (with respect to the index $i$), a piecewise-linear function is formed. Then we can apply existing methods \cite{lipman2010approximating,jimenez2016transforming, amhraoui2022smoothing, arandiga2005interpolation} to find a good approximation, or even an exact function $f$ with the properties as desired. Finally, in the same appendix, we will also discuss other means to prepare the state $\ket{\Phi}$ above, based on \cite{marin2023quantum, nakaji2022approximate}, which also achieve the same complexity $\mathcal{O}\left( \log N\right)$. %Thus, as a side-contribution, we show how to expand the capability of state preparation protocol \cite{mcardle2022quantum} in practice. 

As the next recipe, given the classical description of $\{ \mathscr{D}_r \} $ as above, we have the following lemma, which is crucial to our subsequent construction:
\begin{lemma}[Block-encoding of boundary operator -- Appendix \ref{sec: proofoflemmaentrycomputablematrix} ]
    \label{lemma: entrycomputablematrix}
    Provided the classical knowledge of $ \mathscr{D}_r$ described above, then there is a quantum procedure, using a $\mathcal{O}\left( \log \left(|\Scal_{r-1}||\Scal_r|\right) \right)$-qubits quantum circuit of depth $\mathcal{O}\left( \log \left( |\Scal_{r-1}||\Scal_r|\right) \right)$ that produces an exact block-encoding of $ \frac{\partial_r^\dagger \partial_r}{ |\Scal_{r-1}|\abs{\Scal_r}} $, extra $\mathcal{O}(1)$ ancilla qubit, with further $\mathcal{O}(1)$ classical pre-processing. 
\end{lemma}
This result is originally of \cite{lee2025new} (which is based on \cite{nghiem2025refined}). To provide an overview of key ideas and steps, we note that the classical description of $\mathscr{D}_r$ is translated into the classical description of $\partial_r$ (almost by definition). Given such classical knowledge, using the state preparation algorithm \cite{mcardle2022quantum, marin2023quantum, nakaji2022approximate}, we can prepare the state that contains $\sim \sum_i \partial_r^i \ket{i}$ (where $\partial_r^i$ is the $i$-th column of $\partial_r$, and $\sim$ hides some scaling factor) as a sub-vector. From such the state, a few arithmetic recipes from block-encoding framework \cite{gilyen2019quantum} (e.g., Lemma \ref{lemma: scale}, Lemma \ref{lemma: improveddme} in the Appendix \ref{sec: summaryofnecessarytechniques}) can be applied, resulting in the block-encoding of $ \frac{\partial_r^\dagger \partial_r}{ |\Scal_{r-1}|\abs{\Scal_r}}$. Interested readers can find more details in our Appendix \ref{sec: proofoflemmaentrycomputablematrix}, where we directly quote and discuss the proof from \cite{lee2025new} in detail.

\subsection{Quantum Algorithm}
We use a classical random sampler to draw independent and identically distributed (i.i.d.) elements from $\mathbb{F}_p$. We then form $s\cdot t$ (for $s,t \in \Zbb_+$) vectors $u_1,u_2,...,u_s, v_1, v_2,...,v_t \in \mathbb{F}_p^{|\Scal_r|}$ with randomly i.i.d. entries drawn from $\mathbb{F}_p$. Then we use Lemma \ref{lemma: finitefieldstatepreparation} to prepare the following states $\ket{v_1},\ket{v_2},..., \ket{v_t} ,\ket{u_1},\ket{u_2},...,\ket{u_s}$. The value of $s,t \in \Zbb_+$ is chosen to guarantee a high success probability, as we will see below. 

The reason why we need to prepare the above states with i.i.d. entries is for the purpose of finding the dimension of the rank of $ \Delta_r  $(modulo $p$), which contains the dimension of the homology group of interest $H_r(\mathbb{F}_p)$. Existing quantum algorithms for finding kernel/rank employed in previous works can only handle real or complex field. In the finite field, these methods are no longer guaranteed to produce the correct solution as the arithmetic operation needs to be done modulo $p$. In order to handle the $\rm mod \ p$ operation, we rely on the classical algorithm introduced in \cite{kaltofen1991wiedemann, eberly2017black}. A summary of such algorithms can be found in Algorithm \ref{algo: rankestimatingfinitefield} in the Appendix \ref{sec: randomizedrankestimation}. 

To proceed, via Lemma \ref{lemma: entrycomputablematrix}, we obtain the block-encoding $U_r,U_{r+1}$ of $ \frac{\partial_r^\dagger \partial_r}{ |\Scal_{r-1}|\abs{\Scal_r}}  $ and of  $ \frac{\partial_{r+1}\partial_{r+1}^\dagger }{ |\Scal_{r}|\abs{\Scal_{r+1}}} $, respectively.
%Then we use a combination of Lemma \ref{lemma: scale} and Lemma \ref{lemma: sumencoding} to construct the block-encoding of 
%$$\frac{1}{ \abs{\Scal_r} (|\Scal_{r-1}|+|\Scal_{r+1}|)} \Delta_r. $$ 
%Next, we take the block-encoding of $\frac{1}{|\Scal_{r-1}|\abs{\Scal_r}} \Delta_r$ and act on $\ket{\bf 0}\ket{v_i}$ for any $1 \leq i \leq s$ (and $\ket{\bf 0}$ matches those ancilla qubits required for block-encoding $\frac{1}{|\Scal_{r-1}|\abs{\Scal_r}}  \Delta_r  $). 
Next, we use the Hadamard/SWAP test with advanced amplitude estimation \cite{rall2021faster,rall2023amplitude,aaronson2020quantum} to estimate the overlaps
\begin{align*}
    \bra{\bf 0}\bra{u_j}U_r  \ket{\bf 0}\ket{u_i},    \bra{\bf 0}\bra{u_j}U_{r+1}  \ket{\bf 0}\ket{u_i}
\end{align*}
up to a chosen additive precision $\epsilon$. In particular, as will be shown in the appendix (see Lemma \ref{innerproduct}), the above inner product is equivalent to
$$  \frac{  u_i^\dagger \partial_r^\dagger \partial_r  v_j}{ ||u_i|| \ ||v_j||  \abs{\Scal_r} |\Scal_{r-1}| },  \frac{  u_i^\dagger \partial_{r+1} \partial_{r+1}^\dagger  v_j}{ ||u_i|| \ ||v_j||  \abs{\Scal_r} |\Scal_{r+1}| } $$
%We remark that the entries of $\partial_r ,u_i,v_j $ are integers, as well as the values of $ |\Scal_{r-1}|  |\Scal_r|,|\Scal_{r+1}|$. Therefore, given that the above ratio can be estimated to an additive precision $\epsilon$, we can use the continued fraction algorithm \cite{jones1980continued, hensley2006continued} to find the concrete values of the numerator $ u_i^\dagger \Delta_r v_j$ (and of the denominator as well, but we would not need such value). As the continued fraction algorithm can significantly enhance precision after each iteration, the initial precision $\epsilon$ can be $\mathcal{O}(1)$, e.g., $\frac{1}{2},\frac{1}4$. 
In particular, by choosing
$$\epsilon = \frac{\eta | u_i^\dagger \partial_r^\dagger \partial_r v_j  |}{||u_i|| \ ||v_j|| |\Scal_r||\Scal_{r-1}| }$$
then we can estimate $ u_i^\dagger \partial_r^\dagger \partial_r v_j, u_i^\dagger\partial_{r+1} \partial_{r+1}^\dagger v_j$ and thus, $u_i^\dagger \Delta_r v_j $ to a multiplicative accuracy $\delta$. As $u_i^\dagger \Delta_r v_j $ is ideally an integer, $\eta = 1/2$ suffices to infer the exact estimation, as we simply need to round $u_i^\dagger \Delta_r v_j $ to the nearest integer. 
We then build the matrix $M$ of size $s \times t$ with the entries defined as follows 
\begin{align}
     M_{ij} = u_i^\dagger \Delta_r v_j
\end{align}
Next, we need to take the modulo $p$ of the above values, to obtain $ M_{ij}$ (mod $p$). We then use the classical algorithm in \cite{kaltofen1991wiedemann, eberly2017black} (see algo Algo.~\ref{algo: rankestimatingfinitefield}) to find the rank of $M$ (mod $p$). Denote such rank as $\hat{r}_p$ as the rank of $M$ (mode $p$). As proved in \cite{kaltofen1991wiedemann, eberly2017black}, by choosing $s,t =\mathcal{O}\left( \log_p \frac{1}{\delta} \right)$, the value $\hat{r}_p$ produces the exact rank of $\Delta_r$ with success probability $1-\delta$.  

The last step is straightforward. In order to detect the torsion, we repeat the above algorithm for a few values of $p$, and possibly over real field $\mathbb{R}$. Then we obtain a sequence of values $ \{  \hat{r}_{p_i}\}$ with $\{ p_i\}$ being primes. By comparing them, the torsion can be detected. 

Earlier, we have mentioned that the oracle model can be integrated into our algorithm simply. Here, we elaborate on this. Within the non-oracle model, the classical knowledge/description of the matrices $\{\mathscr{D}_r\}$ allow us to obtain the block-encoding of $ \varpropto  \partial_r^\dagger \partial_r, \partial_{r+1} \partial_{r+1}^\dagger$ (where $\varpropto $ hides the appropriate scaling factor). Within the oracle model, it turns out that these operators can also be block-encoded via the result of \cite{hayakawa2022quantum} (see their Section 6). From such a block-encoding, we can proceed our algorithm outlined above, and thus being able to detect torsion eventually. 

A detailed analysis on the complexity of our algorithm, in both oracle and non-oracle is provided in the Appendix \ref{sec: complexityanalysis}. We summarize our main result in the following:
\begin{theorem}[Torsion Detection]
\label{theorem: torsiondetection}
    Let $K$ be the simplicial complex of interest, having $N$ data points. Then:
    \begin{itemize}
        \item (Non-oracle regime) Provided the classical description of $K$ via the matrices $\{ \mathscr{D}_r\}_{r=1}^N$ as above, there is an algorithm that detects if $K$ has torsion (at order $r$, or of $r$-th homology group) with a failure probability $\delta$. For $S_r^{\max} = \max \{ \Scal_{r-1}^K, \Scal_r^K, \Scal_{r+1}^K \} $, the algorithm has quantum complexity (quantum circuit complexity + total number of iterations)
        \begin{align*}
         %\mathcal{O}\left( \log \left( S_r^{\max} \right)  \frac{1}{\eta}   S_r^{\max}     \right)  \\
            \mathcal{O}\left(  \log \left( S_r^{\max} \right)  \frac{1}{\eta}   S_r^{\max}    \log_p^2 \frac{1}{\delta}  \right) 
        \end{align*}
  and employ a classical procedure of complexity $\mathcal{O}\left(  \log \big( |\Scal_{r-1}|  |\Scal_r||\Scal_{r+1}|\big) +  \log_p^2 \frac{1}{\delta} \right) $. 
 \item (Oracle-regime) Provided the oracle $O_r$ which can verify the inclusion of $r$-simplexes as above, there is an algorithm that detects if $K$ has torsion at order $r$, or of $r$-th homology group) with a failure probability. The algorithm quantum complexity (quantum circuit depth + total number of iterations)
 \begin{align*}
    \mathcal{O}\left(  (N^2+\log |\Scal_r|) N \frac{1}{\eta} \log_p^2 \frac{1}{\delta}\right)
\end{align*}
and employs a classical procedure of complexity $\mathcal{O}\left( \log \big( |\Scal_{r-1}|  |\Scal_r||\Scal_{r+1}|\big)\log^2 \frac{1}{\delta} \right) $.
    \end{itemize}
  %The factor $\kappa_{\Delta_r}$ denotes the condition number of the $r$-th combinatorial Laplacian of $K$. The value $\mathcal{C} \in \Rbb$ above is a randomized factor, with a probabilistic bounded norm as:
   % \begin{itemize}
   %     \item (Weak version) For some $S \in \Rbb_+$ and $S^2 \geq \frac{ |\Scal_r| (p^2-1)}{12}$, the %probability that $\mathcal{C} \leq S$ is:
   %     \begin{align}
   %         \rm Pr( \mathcal{C} \leq S^2 )  \geq 1- \frac{ |\Scal_r|  \left(\frac{p^4}{180} - \frac{p^2}{36} + \frac{1}{45} \right)^2 }{  |\Scal_r|  \left(\frac{p^4}{180} - \frac{p^2}{36} + \frac{1}{45} \right)^2 + (S^2 - \frac{|\Scal_r| (p^2-1)}{12})^2}
   %     \end{align} 
   %     \item (Stronger version) For any $S \in \Rbb_+$:
   %     \begin{align}
   %         \rm Pr( \mathcal{C} \leq S )  \geq  \Phi \left(  \frac{S^2 - \frac{  |\Scal_r| (p^2-1)}{12}}{ \sqrt{ |\Scal_r| }  \left(\frac{p^4}{180} - \frac{p^2}{36} + \frac{1}{45} \right)}\right) - \frac{0.56 (p-1)^6}{2^6  \left(\frac{p^4}{180} - \frac{p^2}{36} + \frac{1}{45} \right)^3 \sqrt{ |\Scal_r| }}
    %    \end{align}
    %\end{itemize}
    %where $\Phi$ is the standard normal cumulative distribution function. 
\end{theorem}

\noindent
\textbf{Potential advantage.} To compare, we point out that the classical algorithm for detecting torsion can still rely on examining the behavior of homology groups $H_r$ over different fields. A naive way is to compute the dimension of kernel of $\Delta_r$ directly, e.g., via exact diagonalization or Gaussian elimination. For a complex having $|\Scal_r|$ $r$-simplexes, we first need to build the boundary operators $\partial_r$, then building the Laplacian $\Delta_r$, resulting in total complexity $\mathcal{O}(|\Scal_r|^\omega)$ where $\omega$ is the matrix multiplication factor. Then applying diagonalization or Gaussian elimination incurs a total complexity $\mathcal{O}\left( |\Scal_r|^3 \right)$. Alternatively, we can use the classical algorithm developed \cite{kaltofen1991wiedemann, eberly2017black} (see Algo.~\ref{algo: rankestimatingfinitefield}). Instead of diagonalizing $\Delta_r$, we perform a sequence of matrix-vector products $v_j^\dagger \Delta_r u_i$ for $i=1,2,...,s$, $j=1,2,...,t$. Then we build a matrix $M$ of size $s \times t$ with $M_{ij} = v_j^\dagger \Delta_r u_i$. The rank, and thus, the kernel of $\Delta_r$ can be found by finding the rank of $M$, which can be done using exact diagonalization or Gaussian elimination. This approach will reduce the time complexity to $\mathcal{O}\left( |\Scal_r|^\omega + \log^2_p \frac{1}{\delta}\right)$, with a success probability $1-\delta$ for $s,t = \mathcal{O}\left( \log \frac{1}{\delta}\right)$. As indicated in the Theorem \ref{theorem: torsiondetection}, quantum algorithm can offer an exponential speed-up with respect to the number of $r$-simplexes $|\Scal_r|$. 

Comparing the classical complexity and our complexity in Theorem \ref{theorem: torsiondetection}, it can be seen that in the non-oracle regime, there is almost a power-of-3 speed-up in $|\Scal_r|$. In the oracle-regime, if $N \ll  \Scal_r$, e.g., when $r$ is sufficiently large and the number of complexes $\Scal_r$ is large (clique-dense), then there is an exponential speed-up in $|\Scal_r|$. On other hand, when $N \sim |\Scal_r|$, then there is no speedup.

\section{Discussion $\&$ Conclusion }
\label{sec: conclusion}
In this work, we have proposed and analyzed a quantum algorithm for detecting torsion. Starting from the mathematical properties, specifically the algebraic structure of the homology groups, we outline a strategy for revealing the torsion from the complex. Our quantum algorithm is built upon severally recent advances in quantum algorithms, including \cite{nghiem2025refined, lee2025new, gilyen2019quantum}, and the classical probabilistic method for rank estimation \cite{kaltofen1991wiedemann, eberly2017black}. As been analyzed, with a certain assist from classical computers and probabilistic guarantee, our quantum algorithm can provide a significant speed-up compared to the classical counterpart. 

As mentioned in the introduction, the potential of quantum computation in TDA has been somewhat limited by the complexity-theoretic result of \cite{schmidhuber2022complexity}. However, the result of this work has revealed that a fundamentally related problem can still be beneficial with quantum computers. Our work has suggested that there should be more room and opportunities for exploring quantum advantage in TDA, beyond the issue of Betti numbers. Below, we list several prospects that we think it is of high value.
\begin{itemize}
    %\item An important factor within our algorithm is the condition number $\kappa_{\Delta_r}$. We point out that in \cite{ubaru2021quantum} (Section 4.3) and \cite{berry2024analyzing} (Section IV), the authors explicitly discussed and constructed some types of complexes that the combinatorial Laplacian $\Delta_r$ admitting low condition number. The complexes provided there are apparently does not have torsion. So, how to generalize these examples to include low condition numbers plus torsion, is of worth pursuing.
    \item Our work has provided a quantum algorithm for detecting torsion. However, the structure of torsion can be complicated, e.g., of the form $\bigoplus_i Z_{r_i} $, and our algorithm cannot determine these factors $\{r_i\}$. The classical approach is to compute the Smith normal form for the boundary, or Laplacian operator. However, this approach is very computationally expensive. A direct translation of this procedure to quantum setting is not known to us. Even if there is, we suspect that the cost is high. Therefore, developing an efficient quantum algorithm for finding these invariant factors is of great interest. 
    \item TDA has been proposed for a while, and the corresponding algorithms have been applied in many contexts. However, from a complexity-theoretic viewpoint, it just has been recently established \cite{crichigno2024clique, schmidhuber2022complexity} that Betti numbers are difficult to estimate, and it is even harder to compute them exactly. Whether if there exists an analogous result regarding the torsion and its structure, specifically the invariant factors $\{r_i\}$, is unclear to us. 
\end{itemize}
We believe that all these questions can foster new development toward quantum topological data analysis. We conclude our work here and leave them for future investigation.

\section*{Acknowledgements}
Part of this work is done when the author is an intern at QuEra Computing Inc. We particularly acknowledge the hospitality of Harvard University where the author has an academic visit during the completion of this project.

\bibliography{ref.bib}
\bibliographystyle{unsrt}

\newpage
\appendix
\section{Block-encoding and quantum singular value transformation}
\label{sec: summaryofnecessarytechniques}
We briefly summarize the essential quantum tools used in our algorithm. For conciseness, we highlight only the main results and omit technical details, which are thoroughly covered in~\cite{gilyen2019quantum}. An identical summary is also presented in~\cite{lee2025new}.

\begin{definition}[Block-encoding unitary, see e.g.~\cite{low2017optimal, low2019hamiltonian, gilyen2019quantum}]
\label{def: blockencode} 
Let $A$ be a Hermitian matrix of size $N \times N$ with operator norm $\norm{A} < 1$. A unitary matrix $U$ is said to be an \emph{exact block encoding} of $A$ if
\begin{align}
    U = \begin{pmatrix}
       A & * \\
       * & * \\
    \end{pmatrix},
\end{align}
where the top-left block of $U$ corresponds to $A$. Equivalently, one can write
\begin{equation}
    U = \ket{\mathbf{0}}\bra{\mathbf{0}} \otimes A + (\cdots),    
\end{equation}
where $\ket{\mathbf{0}}$ denotes an ancillary state used for block encoding, and $(\cdots)$ represents the remaining components orthogonal to $\ket{\mathbf{0}}\bra{\mathbf{0}} \otimes A$. If instead $U$ satisfies
\begin{equation}
    U = \ket{\mathbf{0}}\bra{\mathbf{0}} \otimes \tilde{A} + (\cdots),
\end{equation}
for some $\tilde{A}$ such that $\|\tilde{A} - A\| \leq \epsilon$, then $U$ is called an {$\epsilon$-approximate block encoding} of $A$. Furthermore, the action of $U$ on a state $\ket{\mathbf{0}}\ket{\phi}$ is given by
\begin{align}
    \label{eqn: action}
    U \ket{\mathbf{0}}\ket{\phi} = \ket{\mathbf{0}} A\ket{\phi} + \ket{\mathrm{garbage}},
\end{align}
where $\ket{\mathrm{garbage}}$ is a state orthogonal to $\ket{\mathbf{0}}A\ket{\phi}$. The circuit complexity (e.g., depth) of $U$ is referred to as the {complexity of block encoding $A$}.
\end{definition}

Based on~\cref{def: blockencode}, several properties, though immediate, are of particular importance and are listed below.
\begin{remark}[Properties of block-encoding unitary]
The block-encoding framework has the following immediate consequences:
\begin{enumerate}[label=(\roman*)]
    \item Any unitary $U$ is trivially an exact block encoding of itself.
    \item If $U$ is a block encoding of $A$, then so is $\Ibb_m \otimes U$ for any $m \geq 1$.
    \item The identity matrix $\Ibb_m$ can be trivially block encoded, for example, by $\sigma_z \otimes \Ibb_m$.
\end{enumerate}
\end{remark}

Given a set of block-encoded operators, various arithmetic operations can be done with them. Here, we simply introduce some key operations that are especially relevant to our algorithm, focusing on how they are implemented and their time complexity, without going into proofs. For more detailed explanations, see~\cite{gilyen2019quantum, camps2020approximate}.

\begin{lemma}[Informal, product of block-encoded operators, see e.g.~\cite{gilyen2019quantum}]
\label{lemma: product}
    Given unitary block encodings of two matrices $A_1$ and $A_2$, with respective implementation complexities $T_1$ and $T_2$, there exists an efficient procedure for constructing a unitary block encoding of the product $A_1 A_2$ with complexity $T_1 + T_2$.
\end{lemma}

\begin{lemma}[Informal, tensor product of block-encoded operators, see e.g.~{\cite[Theorem 1]{camps2020approximate}}]\label{lemma: tensorproduct}
    Given unitary block-encodings $\{U_i\}_{i=1}^m$ of multiple operators $\{M_i\}_{i=1}^m$ (assumed to be exact), there exists a procedure that constructs a unitary block-encoding of $\bigotimes_{i=1}^m M_i$ using a single application of each $U_i$ and $\mathcal{O}(1)$ SWAP gates.
\end{lemma}

\begin{lemma}[Informal, linear combination of block-encoded operators, see e.g.~{\cite[Theorem 52]{gilyen2019quantum}}]
    Given the unitary block encoding of multiple operators $\{A_i\}_{i=1}^m$. Then, there is a procedure that produces a unitary block encoding operator of $\sum_{i=1}^m \pm (A_i/m) $ in time complexity $\mathcal{O}(m)$, e.g., using the block encoding of each operator $A_i$ a single time. 
    \label{lemma: sumencoding}
\end{lemma}

\begin{lemma}[Informal, Scaling multiplication of block-encoded operators] 
\label{lemma: scale}
    Given a block encoding of some matrix $A$, as in~\cref{def: blockencode}, the block encoding of $A/p$ where $p > 1$ can be prepared with an extra $\mathcal{O}(1)$ cost.
\end{lemma}

% To show this, we note that the matrix representation of the $R_Y$ rotation gate is given by
% \begin{equation}
%     R_Y(\theta) = \begin{pmatrix}
%         \cos(\theta/2) & -\sin(\theta/2) \\
%         \sin(\theta/2) & \cos(\theta/2) 
%     \end{pmatrix}. 
% \end{equation}

% If we choose $\theta=2\cos^{-1}(1/p)$, then by Lemma~\ref{lemma: tensorproduct}, we can construct a block-encoding of $R_Y(\theta) \otimes \mathbb{I}_{{\rm dim}(A)}$, where ${\rm dim}(A)$ refers to the dimension of the rows (or columns) of the square matrix $A$. This operation results in a diagonal matrix of size ${\rm dim}(A) \times {\rm dim}(A)$ with all diagonal entries equal to $1/p$. Then, by applying Lemma~\ref{lemma: product}, we can construct a block-encoding of
% \begin{equation}
%      \frac{1}{p} \ \mathbb{I}_{{\rm dim}(A)} \cdot A = \frac{A}{p}
% \end{equation}

\begin{lemma}[Matrix inversion, see e.g.~\cite{gilyen2019quantum, childs2017quantum}]\label{lemma: matrixinversion}
Given a block encoding of some matrix $A$  with operator norm $||A|| \leq 1$ and block-encoding complexity $T_A$, then there is a quantum circuit producing an $\epsilon$-approximated block encoding of ${A^{-1}}/{\kappa}$ where $\kappa$ is the conditional number of $A$. The complexity of this quantum circuit is $\mathcal{O}\left( \kappa T_A \log \left({1}/{\epsilon}\right)\right)$. 
\end{lemma}

\begin{lemma}\label{lemma: amp_amp}[\cite{gilyen2019quantum} Theorem 30]
\label{lemma: amplification}
Let $U$, $\Pi$, $\widetilde{\Pi} \in {\rm End}(\mathcal{H}_U)$ be linear operators on $\mathcal{H}_U$ such that $U$ is a unitary, and $\Pi$, $\widetilde{\Pi}$ are orthogonal projectors. 
Let $\gamma>1$ and $\delta,\epsilon \in (0,\frac{1}{2})$. 
Suppose that $\widetilde{\Pi}U\Pi=W \Sigma V^\dagger=\sum_{i}\varsigma_i\ket{w_i}\bra{v_i}$ is a singular value decomposition. 
Then there is an $m= \mathcal{O} \Big(\frac{\gamma}{\delta}
\log \left(\frac{\gamma}{\epsilon} \right)\Big)$ and an efficiently computable $\Phi\in\mathbb{R}^m$ such that
\begin{equation}
\left(\bra{+}\otimes\widetilde{\Pi}_{\leq\frac{1-\delta}{\gamma}}\right)U_\Phi \left(\ket{+}\otimes\Pi_{\leq\frac{1-\delta}{\gamma}}\right)=\sum_{i\colon\varsigma_i\leq \frac{1-\delta}{\gamma} }\tilde{\varsigma}_i\ket{w_i}\bra{v_i} , \text{ where } \Big|\!\Big|\frac{\tilde{\varsigma}_i}{\gamma\varsigma_i}-1 \Big|\!\Big|\leq \epsilon.
\end{equation}
Moreover, $U_\Phi$ can be implemented using a single ancilla qubit with $m$ uses of $U$ and $U^\dagger$, $m$ uses of C$_\Pi$NOT and $m$ uses of C$_{\widetilde{\Pi}}$NOT gates and $m$ single qubit gates.
Here,
\begin{itemize}
\item C$_\Pi$NOT$:=X \otimes \Pi + I \otimes (I - \Pi)$ and a similar definition for C$_{\widetilde{\Pi}}$NOT; see Definition 2 in \cite{gilyen2019quantum},
\item $U_\Phi$: alternating phase modulation sequence; see Definition 15 in \cite{gilyen2019quantum},
\item $\Pi_{\leq \delta}$, $\widetilde{\Pi}_{\leq \delta}$: singular value threshold projectors; see Definition 24 in \cite{gilyen2019quantum}.
\end{itemize}
\end{lemma}
\begin{lemma}
\label{lemma: qsvt}[\cite{gilyen2019quantum} Theorem 56]
\label{lemma: theorem56}  
Suppose that $U$ is an
$(\alpha, a, \epsilon)$-encoding of a Hermitian matrix $A$. (See Definition 43 of~\cite{gilyen2019quantum} for the definition.)
If $P \in \mathbb{R}[x]$ is a degree-$d$ polynomial satisfying that
\begin{itemize}
\item for all $x \in [-1,1]$: $|P(x)| \leq \frac{1}{2}$,
\end{itemize}
then, there is a quantum circuit $\tilde{U}$, which is an $(1,a+2,4d \sqrt{\frac{\epsilon}{\alpha}})$-encoding of $P(A/\alpha)$ and
consists of $d$ applications of $U$ and $U^\dagger$ gates, a single application of controlled-$U$ and $\mathcal{O}((a+1)d)$
other one- and two-qubit gates.
\end{lemma}

%\begin{lemma}[\cite{nghiem2023quantum}]
%\label{lemma: removingfactor}
%    Let $U_A$ be an exact block-encoding of an operator $A/\alpha$, with operator norm of $A/\alpha$, $||A||/\alpha \leq 1$ and $\alpha \geq 1$. Then there is a quantum circuit of depth $\mathcal{O}\left(  \kappa^2 T_A \log^4 \frac{\alpha}{\epsilon} \right) $ that implements the $\epsilon$-approximated block-encoding of $A/\kappa$, where $\kappa$ is the condition number of $A$ and $T_A$ is the circuit complexity of $U_A$. 
%\end{lemma}
%We refer the interested readers to the Appendix F of \cite{nghiem2023quantum} for the proof. Here, we point out and discuss the importance of the above lemma. It can be seen that if $\kappa \leq \alpha$, then to obtain the block-encoding of $A/\kappa$, we can use Lemma \ref{lemma: amp_amp} to multiply the block-encoding of $A/\alpha$ with $\alpha/\kappa$. This way has complexity $\mathcal{O}( \alpha/\kappa)$ which is linear in $\alpha$. At the same time, the above lemma achieves better dependence on $\alpha$, with a trade-off being the dependence on condition number $\kappa$. As we see throughout our work, the factor (similar to $\alpha$) associated with the block-encoding of the boundary operator is usually very high (like of the order of dimension of corresponding matrix), thus the above lemma allows a more efficient way to remove it. 

\section{Abstract Algebra}
\label{sec: reviewalgebra}
In this section, we provide an overview of algebra, showing essential concepts to understand the meaning of torsion. As a consequence, we aim to comprehend the role of coefficients and how it can affect the Betti numbers -- a central problem within TDA, as mentioned in the main text. A more detailed introduction to abstract algebra can be found in standard textbook, e.g., \cite{gallian2021contemporary,grillet2007abstract}.

\begin{definition}[Group]
A group $\mathcal{G}$ is a set equipped with an operation, denoted as $*$, with the following so-called group axioms:
\begin{enumerate}
    \item \textbf{Closure:} if $a,b \in \mathcal{G}$ then $a * b \in X$. 
    \item \textbf{Associativity:} $a*(b*c) = (a*b) *c$.
    \item \textbf{Identity: } There is a namely identity element $e$ such that $a*e = e*a = a$. 
    \item \textbf{Inverse: } Every element $a$ has an inverse element $a^{-1}$ s.t. $a a^{-1} = a^{-1} a = e$. 
\end{enumerate}
\end{definition}

A group is said to be Abelian if the group operation between two elements commute, i.e., for any $a,b \in \mathcal{G}$, we have that $a * b = b*a$. Abelian group is very common in many contexts, e.g., any vector space is an Abelian group. 

\begin{definition}[Torsion]
 Given an Abelian group $\mathcal{G}$, an element $ x \in \mathcal{G}$ is a \textbf{torsion element} if there is some positive integer $n$ that makes $x*x*x*\cdots * x $ ($n$ times) an identity element $e$. If there is no such integer $n$ exists, then $x$ is called \textbf{free}. 
\end{definition}
\noindent
\textbf{Examples:} 
\begin{itemize}
    \item In $\mathbb{Z}_6$ (integers mod 6), which we conveniently denote as $\{ 0, 1,2,3,4,5 \}$, then $2$ is torsion element. 
    \item In $\mathbb{Z}$, there is no torsion except 0 -- the identity element. 
\end{itemize}

Given multiple groups, it is possible to ``build'' a larger group from these smaller groups while keeping their respective structure. The procedure is called \textit{direct sum}. 
\begin{definition}[Direct Sum]
    If $A$ and $B$ are two groups, then their direct sum is:
    \begin{align}
        A \oplus B = \{  (a,b) \ |\  a \in A, b \in B \}
    \end{align}
    with the component-wise operation is defined as:
    \begin{align}
        (a_1,b_1) + (a_2, b_2) = (a_1+ a_2, b_1 + b_2)
    \end{align}
    where $a_1,a_2 \in A, b_1,b_2 \in B$. 
\end{definition}
It can be seen that if $A,B$ is Abelian, then $A \oplus B$ is Abelian. The above construction holds for more composing groups, that is, we can build ``larger'' groups, e.g., $ A_1\oplus A_2\oplus A_3 \oplus \cdots \oplus A_n$.  \\

Given a group $\mathcal{G}$, assumed to be Abelian for simplicity, with group operation $*$. Let $x \in \mathcal{G}$, $n \in \mathbb{Z}$ be some integers, and define $nx \equiv x*x*x*\cdots *x$ ($n$ times). Then $\mathcal{G}$ is said to be finitely generated if every element of $\mathcal{G}$ can be expressed as $ \sum n_i x_i$ where $n_i \in \mathbb{Z}$ and $x_i \in \mathcal{G}$. In this case, $\{x_i\}$ is said to be the generator of $\mathcal{G}$. We point out the following examples to illustrate this definition of generator:
\begin{itemize}
    \item The Abelian group $\mathbb{Z}^2$ is generated by $(1,0)$ and $(0,1)$. 
    \item The group $\mathbb{Z}_6 \equiv \{ 0,1,2,3,4,5\}$ is generated by $1$.  
\end{itemize}

At the heart of abstract algebra is the following theorem, popularly known as the structured theorem: 
\begin{theorem}[Structured Theorem for Finitely Generated Abelian Groups]
\label{thm: structuredtheorem}
    Every finitely generated Abelian group $\mathcal{G}$ can be decomposed as:
    \begin{align}
        \mathcal{G} \cong \Zbb^r \oplus \Zbb_{d_1} \oplus \Zbb_{d_2} \oplus \cdots \oplus \Zbb_{d_m}
    \end{align}
    where:
    \begin{enumerate}
        \item $r$ = rank, which is the number of independent ``infinite'' directions. 
        \item Each $\Zbb_{d_i} $ is a finite cyclic group of size $d_i$. 
        \item It holds that $d_1 | d_2 | \cdots | d_m$ where $d_i | d_{i+1}$ denotes the divisibility. 
    \end{enumerate}
     The first part $\Zbb^r$ is called \textit{free} part and $\Zbb_{d_1} \oplus \Zbb_{d_2} \oplus \cdots \oplus \Zbb_{d_m} $ is called \textit{torsion} part.
\end{theorem}
Aside from group, there are other algebraic objects, including ring and field. 
\begin{definition}[Ring]
    A ring $R$ is a set equipped with two operations: 
    \begin{enumerate}
        \item Addition (+), making $(R,+)$ an Abelian group. 
        \item Multiplication $(*)$, making $(R, *)$ a semingroup with an associativity $(a*b)*c = a*(b*c)$. 
        \item Distributivity: for all $a,b,c \in R$, it holds that:
        \begin{align}
            a*(b+c) = a*b + a*c \\
            (a+b)*c = a*c + b*c
        \end{align}
    \end{enumerate}
    A ring might not have a multiplicative identity. If it does, denote the identity as $1_R$ (which is not the identity element of the Abelian group under (+) operation), then we call $R$ a ring with unity. If the multiplicative operation commutes, for example, $a*b = b*a$, then the ring $R$ is called a commute ring (possibly with unity). \\

    \noindent
    \textbf{Example:} $\Zbb$ (the integers) is a commutative ring with unity.
\end{definition}

\begin{definition}[Field]
    A field is a commutative ring $R$ with unity, where it holds that every element has a multiplicative inverse, e.g., for any $a \in R$, there is another $b$ such that $a*b = 1_R$. \\
    
    \noindent
    \textbf{Examples:} $\mathbb{Q}, \mathbb{R},\mathbb{C}$ are infinite fields. $\mathbb{Z}_p$ with $p$ prime is a field. 
\end{definition}

In the above, we have defined what it means to be a group and an Abelian group. It can be seen that a vector space is an Abelian group, with vector addition $(+)$ as a group operation. A vector space is usually defined over some field $\mathbb{F}$, and a vector space is also equipped with a scalar multiplication (with a few axioms associated). A generalization of vector space is called a \textit{module}, which is defined over a ring instead of a field. 
\begin{definition}[Module]
    Let $R$ be some ring. An $R$-module $M$ is:
    \begin{enumerate}
        \item An Abelian group $(M,+)$.
        \item Equipped with scalar multiplication $R \times M \longrightarrow M$, written $(r,m) \longrightarrow rm$ satisfying: 
        \begin{itemize}
            \item $r(m_1+ m_2) = r m_1 + rm_2$. 
            \item $(r_1+r_2) m = r_1  m + r_2 m  $. 
            \item $(r_1 r_2) m = r_1( r_2 m)$. 
            \item $1_R m = m$ (where $1_R$ denotes the identity element in the ring $R$). 
        \end{itemize}
    \end{enumerate}
\end{definition}
We recall a basis of a vector space in which an arbitrary element of some space can be written as a linear combination of the elements in the basis. For a set $S$, the free module $F(S)$ is made of (formal) linear combination of elements of $S$ with coefficients from a ring $R$. In linear algebra, given two vector spaces, one can construct the tensor product of them. In the setting of module, two modules can be tensor producted. Yet, the rule is different.
\begin{definition}[Tensor product of module]
    Given two $R$-modules $M,N$, the tensor product $M \otimes_R N$ is defined as:
    \begin{align}
        M \otimes_R N = F / \text{(bilinear relation)}
    \end{align}
    where $F$ is the free module generated by the pair $(m,n)$ and the bilinear relation is: 
    \begin{itemize}
        \item $(m_1+ m_2, n) = (m_1,n) + (m_2,n)$. 
        \item $(m,n_1+ n_2) = (m, n_1) + (m,n_2)$. 
        \item $(rm,n) = r (m,n) =(m, rn)$. 
    \end{itemize}
   The following properties are imposed within tensor product: 
    \begin{enumerate}
        \item $(m_1+ m_2) \otimes n = m_1 \otimes n + m_2 \otimes n$. 
        \item $ m \otimes (n_1 + n_2) = m \otimes n_1 + m \otimes n_2$. 
        \item $(rm)\otimes n = m \otimes (rn)$. 
    \end{enumerate}
\end{definition}
It can be seen that any Abelian group $\mathcal{G}$, with group operation $(*)$, is a $\Zbb$-module, by defining the action $n x = x * x * x *\cdots *x $ (n times) for $n \in \Zbb$ and $x \in \mathcal{G}$. We point out a few properties regarding computing tensor product between $\Zbb$-modules and other module:
\begin{lemma}[Tensor product of $\Zbb$-modules]
\label{lemma: propertytensorproductmodule}
Let $\Zbb$ be the set of integers. Then it holds that:
    \begin{itemize}
    \item $\Zbb^r \otimes_{\Zbb} N \cong N^r$ for arbitrary $\Zbb$-module $N$.
    \item $\Zbb_m \otimes_{\Zbb} \Zbb_n \cong \Zbb_{\rm gcd (m,n)} $.
\item $\Zbb_m \otimes_{\Zbb} Q = 0$ for $Q$ being a field of characteristic zero. 
\item $\Zbb^r \otimes_{\Zbb} Q \cong Q^r $
\end{itemize}
\end{lemma}
For illustration, we provide the following examples. \\

\noindent
\textbf{Example 1.} Let $M = \Zbb^2 \oplus \Zbb_4$, and $N = \Zbb_6$ be $\Zbb$-modules. Then we have the following. 
\begin{align}
    M \otimes_{\Zbb} N &= (\Zbb^2 \oplus \Zbb_4) \otimes_{\Zbb} \Zbb_6 \\
    &= ( \Zbb^2 \otimes_{\Zbb} \Zbb_6) \oplus (\Zbb_4  \otimes_{\Zbb} \Zbb_6 )\\
    &= \Zbb_6^2 \oplus \Zbb_2
\end{align}

\noindent
\textbf{Example 2.} Let $M = \Zbb^2 \oplus \Zbb_4$ and $N = Q$. Then we have: 
\begin{align}
    M \otimes_{\Zbb} N &= ( \Zbb^2 \oplus \Zbb_4 ) \otimes_{\Zbb} Q \\
    &= ( \Zbb^2 \otimes_{\Zbb} Q) \oplus (\Zbb_4 \otimes_{\Zbb} Q) \\
    &= 0 
\end{align}
From the above example, we see that whenever a $\Zbb$-module is tensor producted with a field (of characteristic zero), then the resulting module is zero. Subsequently, we will point out that this accounts for the fact that homology built on real coefficients miss the torsion part. 

\section{Algebraic Topology}
\label{sec: reviewofalgebraictopology}

This appendix provides a concise overview of the fundamental concepts in algebraic topology relevant to our work. We primarily follow the exposition of Nakahara~\cite{nakahara2018geometry, hatcher2005algebraic}, to which we refer interested readers for comprehensive treatment of the subject.

\begin{definition}[Simplex]
\label{def:simplex}
Let $p_0, p_1, \ldots, p_r \in \mathbb{R}^m$ be geometrically independent points where $m \geq r$. The $r$-simplex $\sigma_r = [p_0, p_1, \ldots, p_r]$ is defined as:
\begin{equation}
\sigma_r = \left\{ x \in \mathbb{R}^m : x = \sum_{i=0}^r c_i p_i, \quad c_i \geq 0, \quad \sum_{i=0}^r c_i = 1 \right\},
\end{equation}
where the coefficients $(c_0, c_1, \ldots, c_r)$ are called the barycentric coordinates of $x$.
\end{definition}

Geometrically, a 0-simplex $[p_0]$ represents a point, a 1-simplex $[p_0, p_1]$ represents a line segment, a 2-simplex $[p_0, p_1, p_2]$ represents a triangle, and higher-dimensional simplices generalize this pattern to higher dimensions.

An $r$-simplex can be assigned an orientation. For instance, the 1-simplex $[p_0, p_1]$ has orientation $p_0 \to p_1$, which differs from $[p_1, p_0]$. Throughout this work, we adopt the convention that for an $r$-simplex $[p_0, p_1, \ldots, p_r]$, the indices are ordered from low to high, indicating the canonical orientation.

\begin{definition}[Face and simplicial complex]
For an $r$-simplex $[p_0, p_1, \ldots, p_r]$, any $(s+1)$-subset of its vertices defines an $s$-face $\sigma_s$ where $s \leq r$. A simplicial complex $K$ is a finite collection of simplices satisfying:
\begin{enumerate}
\item Every face of a simplex in $K$ is also in $K$
\item The intersection of any two simplices in $K$ is either empty or a common face of both simplices
\end{enumerate}
The dimension of $K$ is $\dim(K) = \max\{r : \sigma_r \in K\}$.
\end{definition}

\begin{definition}[Chain group]
\label{def: chaingroup}
Let $K$ be an $n$-dimensional simplicial complex. The $r$-th chain group $C_r^K$ is the free abelian group generated by the oriented $r$-simplices of $K$. For $r > \dim(K)$, we define $C_r^K = 0$. Formally, let $S_r^K = \{\sigma_{r,1}, \sigma_{r,2}, \ldots, \sigma_{r,|S_r^K|}\}$ denote the set of $r$-simplices in $K$. An $r$-chain is an element of the form:
\begin{equation}
c_r = \sum_{i=1}^{|S_r^K|} c_i \sigma_{r,i}
\end{equation}
where $c_i \in \mathbb{R}$ are real coefficients. The group operation is defined by:
\begin{equation}
c_r^{(1)} + c_r^{(2)} = \sum_{i=1}^{|S_r^K|} \left( c_i^{(1)} + c_i^{(2)} \right) \sigma_{r,i},
\end{equation}
making $C_r^K$ a free abelian group of rank $|S_r^K|$.
\end{definition}
In fact, since the coefficients $\{ c_i \}$ belong to a field $\Rbb$, the group $C_r^K$ is also a vector space. In the following, we use the Abelian group/vector space interchangeably. The boundary operator $\partial_r : C_r^K \to C_{r-1}^K$ is fundamental to homological algebra.

\begin{definition}[Boundary operator]
For an $r$-simplex $[p_0, p_1, \ldots, p_r]$, the boundary operator is defined as:
\begin{equation}
\partial_r [p_0, p_1, \ldots, p_r] = \sum_{i=0}^r (-1)^i [p_0, p_1, \ldots, \hat{p_i}, \ldots, p_r],
\end{equation}
where $\hat{p_i}$ indicates that vertex $p_i$ is omitted, yielding an $(r{-}1)$-simplex.

For an $r$-chain $c_r = \sum_{i=1}^{|S_r^K|} c_i \sigma_{r,i}$, we extend linearly:
\begin{equation}
\partial_r c_r = \sum_{i=1}^{|S_r^K|} c_i \partial_r \sigma_{r,i}.
\end{equation}
\end{definition}

The boundary operators form a chain complex:
\begin{equation}
0 \xrightarrow{} C_n^K \xrightarrow{\partial_n} C_{n-1}^K \xrightarrow{\partial_{n-1}} \cdots \xrightarrow{\partial_1} C_0^K \xrightarrow{\partial_0} 0.
\end{equation}

\begin{definition}[Cycles, boundaries, and homology]
An $r$-chain $c_r$ is called an $r$-cycle if $\partial_r c_r = 0$. The collection of all $r$-cycles forms the $r$-cycle group $Z_r^K = \ker(\partial_r)$. Conversely, an $r$-chain $c_r$ is called an $r$-boundary if there exists an $(r{+}1)$-chain $d_{r+1}$ such that $\partial_{r+1} d_{r+1} = c_r$. The set of all $r$-boundaries forms the $r$-boundary group $B_r^K = \textnormal{im}(\partial_{r+1})$.
\end{definition}
A fundamental property of boundary operators is that $\partial_r \circ \partial_{r+1} = 0$, which ensures that every boundary is also a cycle, i.e., $B_r^K \subseteq Z_r^K$. This inclusion allows us to define the $r$-th homology group/space as the quotient group/space $H_r^K = Z_r^K / B_r^K$, which captures the notion of cycles that are not boundaries.

\begin{definition}[Betti numbers]
The $r$-th Betti number is defined as:
\begin{equation}
\beta_r = \textnormal{dim}(H_r^K) = \textnormal{dim}(Z_r^K) - \textnormal{dim}(B_r^K).
\end{equation}
\end{definition}
In the group-theoretic language, the above dimension, e.g., $\textnormal{dim}(H_r^K) $, is replaced by $\rm rank (H_r^K)$. As we mentioned, we use the notion of group/space interchangeably. For computational purposes, we can utilize the combinatorial Laplacian:
\begin{equation}
\Delta_r = \partial_{r+1} \partial_{r+1}^\dagger + \partial_r^\dagger \partial_r,
\end{equation}
where $\partial_r^\dagger$ denotes the adjoint of $\partial_r$. A fundamental result in algebraic topology establishes that:
\begin{equation}
H_r^K \cong \ker(\Delta_r),
\end{equation}
providing a direct method for computing Betti numbers via kernel dimension.

A central theorem in algebraic topology states that homology groups constitute topological invariants~\cite{hatcher2005algebraic}:

\begin{theorem}[Topological invariance of homology]
If two topological spaces $X$ and $Y$ are homeomorphic, then their homology groups are isomorphic: $H_r^X \cong H_r^Y$ for all $r \geq 0$. Consequently, their Betti numbers are equal: $\beta_r^X = \beta_r^Y$.
\end{theorem}
This invariance property makes Betti numbers powerful tools for topological classification and forms the mathematical foundation for their applications in topological data analysis.

\section{Role of Coefficients}
\label{sec: roleofcoefficients}
In the above, we have built the chain group/space by taking (formal) linear combination of simplexes, with coefficients coming from a field. More concretely, we have defined the $r$-chain as: 
\begin{equation}
c_r = \sum_{i=1}^{|S_r^K|} c_i \sigma_{r,i}
\end{equation}
where $\{c_i\} \in \Rbb$. Instead, if we choose $\{c_i\}$ from the set of integers $\Zbb$, then the resulting set $C_r^K \equiv \{c_r = \sum_{i=1}^{|S_r^K|} c_i \sigma_{r,i}  \} $ is not a vector space, but just an Abelian group. We emphasize the subtlety that a vector space is inherently an Abelian group, but an Abelian group is not necessarily a vector space (as the multiplication by a scalar might not be properly defined, and also the scalar does not necessarily come from a field).  

Changing the coefficients results in the change in the algebraic structure of homology groups. Suppose that we begin with the coefficients $\{c_i\} \in \Zbb$, then the $k$-th homology group $H_r(\Zbb)$ is an Abelian group (where we specify $\Zbb$ to explicitly imply the choice of coefficients from $\Zbb$). According to Theorem \ref{thm: structuredtheorem}, $H_r(\Zbb)$ admits the following decomposition:
\begin{align}
    H_r(\Zbb) \cong  \rm Free \big( H_r(\Zbb) \big) \oplus \rm Torsion\big( H_r(\Zbb)  \big)
\end{align}
where the free part $\rm Free$ generally has the form $\Zbb^r$ in which $r$ is called the rank, or the Betti numbers. The torsion part $\rm Tor$ is generally of the form $\Zbb_{d_1} \oplus \Zbb_{d_2} \oplus \cdots $ with the divisibility relation $d_i | d_{i+1}$. We can see that the $\rm Tor$ part is the main difference between homology with integer coefficients and real coefficients. To be more specific, if we change the choice of coefficients $\{c_i\}$ from $\Zbb $ to $\Rbb$, then there is a so-called universal coefficient theorem, which relates the change in the structure of homology group (under corresponding coefficients' type):
\begin{align}
    H_r(\Rbb) \cong H_r(\Zbb) \otimes_{\Zbb} \Rbb  \oplus \rm Tor \left( H_{r-1} (\Zbb), \Rbb   \right)
\end{align}
where the last term $\rm Tor \left( H_{r-1} (\Zbb), \Rbb   \right) $ is the \textit{torsion product functor}, which is defined as:
\begin{definition}
    [Torsion product functor]
    Let $\Zbb$ be the set of integers, and $m$ be some integers. Then for a cyclic group $\Zbb_m$ and another $\Zbb$-module $R$, $\rm Tor (\Zbb_m,  R) \cong [ r \in R | \  m r = 0 ] $. In other words, it is the $n$-torsion subgroup of $R$. 
\end{definition}
We remark that this $\rm Tor$ is different from the $\rm Torsion$ that we used earlier. For convenience, we point out the following useful properties associated with $\rm Tor$, and refer the readers to standard textbook for derivation. 
\begin{lemma}[$\rm Tor$ of $\Zbb$-modules]
\label{lemma: torsionfunctor}
Let $\Zbb$ be the set of integers, $p$ be a prime number and $R$ be some $\Zbb$-module.
\begin{itemize}
\item $\rm Tor (\Zbb, R) = 0$
    \item  $  \rm Tor \left( \bigoplus_i \Zbb_{m_i} , R \right)  = \bigoplus_i \rm Tor \left( \Zbb_{m_i}, R\right)$.
    \item Let $\mathbb{F}_p \equiv \Zbb/p\Zbb$ be the finite field of size $p$. Then:
    \begin{align}
    \rm Tor \left( \Zbb_m, \mathbb{F}_p \right)  \cong \begin{cases}
        \mathbb{F}_p \text{\ if p $\mid$ n} \\
        0  \text{\ if p $\nmid$ n} \\
    \end{cases}
\end{align}
\item For arbitrary $m$ (not necessarily a prime), it holds that:
\begin{align}
    \rm Tor (\Zbb_n, \Zbb_m) \cong \Zbb_{\rm gcd(m,n)}
\end{align}
\item For $R = \mathbb{Q}$ -- a divisible group, $ \rm Tor \left( \Zbb_{m}, R\right) = 0$ for any finite $m$. 
\end{itemize}
\end{lemma}
\noindent
Thus, a direct use of the above lemma leads us to:
\begin{align}
     H_r(\Rbb) \cong H_r(\Zbb) \otimes_{\Zbb} \Rbb 
\end{align}
To proceed, we consider the term $H_r(\Zbb) \otimes_{\Zbb} \Rbb $, which is:
\begin{align}
    H_r(\Zbb) \otimes_{\Zbb} \Rbb &\cong \left( \rm Free \big( H_r(\Zbb) \big) \oplus \rm Torsion\big( H_r(\Zbb)  \big) \right) \otimes_{\Zbb} \Rbb \\
    &= \left(  \rm Free \big( H_r(\Zbb) \big) \otimes_{\Zbb} \Rbb \right) \oplus \left(\rm Torsion\big( H_r(\Zbb)  \big) \otimes_{\Zbb} \Rbb  \right) 
\end{align}
We recall the properties from Lemma \ref{lemma: propertytensorproductmodule} that, for any integers $m$, $\Zbb_m \otimes_{\Zbb} Q = 0$ for any field $Q$ of characteristic zero. At the same time, the torsion part $ \rm Torsion\big( H_r(\Zbb)  \big)$ is $\cong \Zbb_{d_1} \oplus \Zbb_{d_2} \oplus \cdots$, which results in 
\begin{align}
    \left(\rm Torsion\big( H_r(\Zbb)  \big) \otimes_{\Zbb} \Rbb  \right)  = 0
\end{align}
For the first part $  \rm Free \big( H_r(\Zbb) \big) \otimes_{\Zbb} \Rbb $, we can use the first property of Lemma \ref{lemma: propertytensorproductmodule}, and also $\rm Free \big( H_r(\Zbb) \big) \cong \Zbb^r $, which results in:
\begin{align}
      \rm Free \big( H_r(\Zbb) \big) \otimes_{\Zbb} \Rbb  \cong \Rbb^r
\end{align}
which implies that 
\begin{align}
    H_r(\Rbb) \cong \Rbb^r
\end{align}
suggesting that $H_r(\Rbb)$ behaves like a vector space. Thus, the above deduction reveals how the choice of coefficients influences the algebraic structure of homology, and that the typical choice of coefficients from $\Rbb$ or $\Zbb_2$ may hide the torsion part.

\section{A few examples}
\label{sec: afewexamples}
To make the role of torsion clearer, we provide a few concrete examples showing how the torsion is intrinsic to topological space. We particularly provide specific topological spaces with their corresponding homology groups. We organize these examples into two categories; one contains torsion-free topological space, while the other contains spaces with torsion.\\

\noindent
\textbf{Torsion-free:}
\begin{itemize}
    \item Sphere $S^n$: $H_0(\Zbb) \cong \Zbb$, $H_n(\Zbb) \cong \Zbb$. For any $0\leq m < n$, $H_m(\Zbb) \cong 0$. 
    \item Torus $T^n$: $H_r(T^n, \Zbb) \cong \Zbb^{ \binom{n}{k}}$. 
    \item Closed orientable surfaces (with genus $\geq 1$): $H_0(\Zbb) \cong \Zbb, H_1 (\Zbb)\cong \Zbb{2g}$, $H_2(\Zbb) \cong \Zbb$. 
    \item Complex projective space $\mathbb{CP}^n$: $H_{2k}(\Zbb) \cong \Zbb$ for $0 \leq k <n$; $H_{2k+1}(\Zbb) \cong 0$.
    \item Quaternionic projective space $\mathbb{HP}^n$: $H_{4k}(\Zbb) \cong \Zbb$ for $0 \leq k < n$. Otherwise 0. 
\end{itemize}

\noindent
\textbf{Torsion: } 
\begin{itemize}
    \item Real projective space $\mathbb{RP}^n$: $H_0(\Zbb) \cong \Zbb$. For $1 \leq i \leq n-1$, $H_i(\Zbb) \cong \Zbb_2$ if even $i$, and $H_i (\Zbb) \cong 0$ for odd $i$. $H_n(\Zbb) \cong \Zbb$ if $n$ is odd, and $H_n(\Zbb) \cong 0$ if $n$ is even. 
    \item Lens space $L(p,q)$ (3-manifolds): $ H_0(\Zbb) \cong \Zbb$, $H_1(\Zbb) \cong \Zbb_p$,  $H_2(\Zbb) \cong 0, H_3(\Zbb) \cong \Zbb$. 
    \item Klein bottle $K$: $H_0(\Zbb) \cong \Zbb, H_1(\Zbb) \cong \Zbb \oplus \Zbb_2, H_2(\Zbb) 0$. 
    \item Non-orientable close surface $N_g$ (connected sum of $g$ copies of $\mathbb{RP}^2$):
    $H_0(\Zbb) \cong 0, H_1(\Zbb) \cong \Zbb^{g-1} \oplus \Zbb_2, H_2(\Zbb) = 0$. 
    \item Moorse space $M(\Zbb_m, n)$: $H_n(\Zbb) \cong \Zbb_m$. 
\end{itemize}

\section{Key insight}
\label{sec: keyinsight}
In the following, we describe the insight, which is a continuation of the previous section, that underlies our subsequently main algorithm to detect the possible torsion of a given simplicial complex, denoted $K$. Throughout the following, we omit this simplex notation $K$ and naturally impose this setting. We recall from the previous section the universal coefficient theorem:
\begin{align}
    H_r(\Rbb) \cong H_r(\Zbb) \otimes_{\Zbb} \Rbb  \oplus \rm Tor \left( H_{r-1} (\Zbb) , \Rbb  \right)
\end{align}
which reflects the change in algebraic structure upon a change in the coefficients from $\Zbb $ to $\Rbb$. If instead of $\Rbb$, we choose another ring/field $Q$, then the formula still holds:
\begin{align}
    H_r(Q) \cong H_r(\Zbb) \otimes_{\Zbb} Q  \oplus \rm Tor \left( H_{r-1} (\Zbb), Q  \right)
\end{align}
According to Theorem \ref{thm: structuredtheorem}, $H_r(\Zbb)$ can be decomposed as:
\begin{align}
    H_r(\Zbb) \cong \Zbb^{\beta_r} \oplus \left( \bigoplus_i Z_{r_i} \right)
\end{align}
Then we have that the above formula is:
\begin{align}
    H_r(Q ) \cong  \left( \Zbb^{\beta_r} \otimes_{\Zbb} Q  \right) \oplus \left( \bigoplus_i  Z_{r_i} \otimes_{ \Zbb} Q \right) \oplus  \rm Tor \left( H_{r-1} (\Zbb), Q \right)
\end{align}
To make the above group become a vector space (since linear algebra is a more natural language in quantum computation), we choose $Q$ to be some field. For a reason that will be clear later, we choose $Q = \mathbb{F}_p \equiv \Zbb/p\Zbb$ for $p$ being a prime. Then for $Q= \mathbb{F}_p$,  the above equation can be written as:
\begin{align}
    H_r(\mathbb{F}_p) \cong  \left( \Zbb^{\beta_r} \otimes_{\Zbb} \mathbb{F}_p  \right) \oplus \left( \bigoplus_i  \Zbb_{r_i} \otimes_{ \Zbb} \mathbb{F}_p \right) \oplus  \rm Tor \left( H_{r-1} (\Zbb), \mathbb{F}_p  \right)
\end{align}
Again, by Theorem \ref{thm: structuredtheorem}, we have:
\begin{align}
    H_{r-1} (\Zbb) \cong  \Zbb^{\beta_q}  \oplus \left( \bigoplus_i Z_{q_i} \right)
\end{align}
Via Lemma \ref{lemma: torsionfunctor}, we have that:
\begin{align}
    \rm Tor\left( H_{r-1} (\Zbb) , \mathbb{F}_p \right) & \cong \rm Tor \left(\Zbb^{\beta_q}  \oplus \left( \bigoplus_i Z_{q_i} \right), \mathbb{F}_p \right)  \\
    & \cong \rm Tor \left( \left( \bigoplus_i Z_{q_i} \right), \mathbb{F}_p \right)  \\
    & \cong \bigoplus_{i, q_i \mid p} \mathbb{F}_p
\end{align} 
By Lemma \ref{lemma: tensorproduct}, and also the fact that $\mathbb{F}_p \equiv \Zbb_p$, we have that:
\begin{align}
     \left( \bigoplus_i  \Zbb_{k_i} \otimes_{ \Zbb} \mathbb{F}_p \right) &\cong \bigoplus_{i, k_i \mid p} \mathbb{F}_p \\
      \left( \Zbb^{\beta_r} \otimes_{\Zbb} \mathbb{F}_p  \right)  &\cong \mathbb{F}_p^{\beta_r}
\end{align}
Gathering everything, we have:
\begin{align}
    H_r(\mathbb{F}_p) \cong  \mathbb{F}_p^{\beta_r} \oplus \left(\bigoplus_{i, r_i \mid p} \mathbb{F}_p \right) \oplus \left(\bigoplus_{i, q_i \mid p} \mathbb{F}_p \right) 
\end{align}
We recall from the previous section that if we choose the field $Q \equiv \Rbb$, then:
\begin{align}
    H_r(\Rbb) \cong \Rbb^{\beta_r}
\end{align}
It means that, the rank of the $k$-th homology space over $\mathbb{F}_p$ field contains additional factors from the torsion part, i.e.,:
\begin{align}
\rm dim \left( H_r(\mathbb{F}_p) \right) = \beta_r + t_r + t_{r-1} 
\end{align}
where $t_r, t_{r-1}$ is the sum of the cyclic summands (of $k$-th homology group/space and $(k-1)$-th homology group/space, respectively) of order divisible by a prime $p$. We recall from the previous section \ref{sec: roleofcoefficients} that: 
\begin{align}
    H_r(\Rbb) \cong \Rbb^{\beta_r} 
\end{align}
which implies 
\begin{align}
    \dim \left(  H_r(\Rbb)\right)  = \beta_r
\end{align}
Our strategy is built on this, as we proceed to estimate the dimension of $k$-th homology group over two different fields $\Rbb$ and $\mathbb{F}_p$, and then compare them. If their dimensions are equal, then it indicates that there is no torsion. Otherwise, there is torsion. The problem then is, what value of $p$ should we choose ? We recall that $p$ is a prime, and it is of a known fact that arbitrary integers can be decomposed as products of different primes. Thus, in general, without knowing the order of summands in advance (which is apparent, because if we know the order already, then the torsion is already known), the only strategy is to try as many primes as possible, e.g., $p=2,3,5 ..., 7$, etc. In the following section, we construct a quantum algorithm to execute the estimation of the desired dimension, and subsequently, the comparison. 

\section{Quantum Algorithm}
\label{sec: quantumalgorithm}

\subsection{Input model}
As seen from the main text, our input model being similar to the recent work \cite{lee2025new}, namely, the non-oracle model. Subsequently, we will also discuss how the oracle model can be used within our algorithm. Let $\Scal_r = \{ \sigma_{r_1}, \sigma_{r_2}, ... \}$ denotes the set of $r$-simplexes in the complex of interest $K$. For some $r$, let $\mathscr{D}_r$ denotes the matrix of size $ |\Scal_{r-1}| \times |\Scal_r|$, which encodes the structure of the given complex as follows. For a given column index, say $j$ (with $1 \leq j \leq |\Scal_r|$), we have:
\begin{align}
    (\mathscr{D}_r)_{ij} = \begin{cases}
        1 \text{ \ if $\sigma_{r_i} \cap \sigma_{(r-1)_j} \neq \emptyset$  } \\
        0 \text{ \ if $\sigma_{r_i} \cap \sigma_{(r-1)_j} = \emptyset$  } 
    \end{cases}
\end{align}
Intuitively, the matrix $\mathscr{D}_r$ encodes the local, mutual relations between all $r$-simplexes and $(r-1)$-simplexes. In addition, as a $r$-simplex shares $r+1$ faces with other $(r-1)$-simplexes, then for a given column, there are $(r+1)$ nonzero entries. The assumption we make in our work is that the entries of $\{ \mathcal{D}_r\}$ are classically known. We remark that the classically known entries is in the sense, or perspective of Lemma \ref{lemma: efficientstatepreparation}, where the entries can be efficiently loaded into a (large) quantum state. 

As also discussed in \cite{lee2025new}, whether this model arises in practical scenario is unknown. However, in reality, we can achieve this model by directly specifying the simplexes. The method of specifying simplexes was also suggested in \cite{schmidhuber2022complexity} as a mean to overcome a high-complexity subroutine required in the original LGZ algorithm. For instance, let $K$ be the complex of interest and $v_0,v_1,....,v_N$ be the $N$ data points, or vertices in $K$. If we pick:
\begin{align}
   \Scal_3 =  [v_0,v_1,v_3], [v_2,v_3,v_7], [v_8,v_9, v_{16}],
\end{align}
as 3-simplexes $\in K$, and:
\begin{align}
   \Scal_2 = [v_0,v_1], [v_1,v_3], [v_0,v_3], [v_2,v_3], [v_3,v_7], [v_2,v_7], [v_8,v_9] [v_8,v_{16}],  [v_9,v_{16}], [v_5,v_7], [v_6,v_8]
\end{align}
as 2-simplexes in $\in K$. Then we organize the above information as a matrix of size  $|\Scal_2|\times |\Scal_3|$, resulting in the matrix $\mathscr{D}_3$. The same construction holds for different orders, resulting in the matrices $\{\mathscr{D}_r\}$. In general, we can reformulate our problem as follows.
\begin{center}
    \textit{Given the classical knowledge of the matrix $\mathscr{D}_r$ of size $|\Scal_{r-1}| \times |\Scal_r| $, with the property that any column has $r+1$ nonzero entries being 1. At the order $r$, does the simplicial complex $K$ with structured encoded in this matrix have torsion ?}
\end{center}

\subsection{Procedure}
Earlier, we have mentioned the strategy for detecting torsion, which is based on comparing the dimension of $r$-th homology space/group $H_k$ over two fields, $\Rbb$ and $\mathbb{F}_p$. Over $\Rbb$, the dimension of $H_k(\Rbb)$ is in fact the $r$-th Betti number commonly used in the literature. Yet, over $\mathbb{F}_p$, the dimension of $H_k(\mathbb{F}_p)$ can change, depending on whether the complex of interest has torsion. In the following, we discuss in detail the quantum algorithm for estimating the dimension of $H_k$ over $\Rbb$ and $\mathbb{F}_p$. Before that, we provoke the following helpful lemma:
\begin{lemma}
\label{innerproduct}
    Let $U_A$ be a unitary block-encoding of a matrix $A \in \mathbb{C}^{N \times N}$ as in Definition \ref{def: blockencode}. Let $\ket{\phi_1} = \frac{1}{||\phi_1||}\phi_1, \ket{\phi_2} = \frac{1}{||\phi_2||} \phi_2$ two quantum states of dimension $N$ with a known preparation procedure. Then it holds that: 
    \begin{align}
        \bra{\bf 0}\bra{\phi_2} U_A \ket{\bf 0}\ket{\phi_1} =   \bra{\phi_2} A \ket{\phi_1}= \frac{1}{||\phi_1|| \ ||\phi_2||} \phi_2^\dagger A \phi_1
    \end{align}
    which can be estimated to some additive or multiplicative accuracy via Hadamard/SWAP test primitive. 
\end{lemma}
To show this, we first observe that, according to Def \ref{def: blockencode}, the action of $ U_A$ on $\ket{\bf 0}\ket{\phi_1}$ is $U_A \ket{\bf 0}\ket{\phi_2} = \ket{\bf 0} A \ket{\phi_2} + \ket{\rm Garbage}$ where $\ket{\rm Garbage}$ is orthogonal to $ \ket{\bf 0} \ket{\alpha} $ for any $\alpha$. Then by taking inner product: 
\begin{align}
\begin{split}
    \bra{\bf 0}\bra{\phi_2} U_A \ket{\bf 0}\ket{\phi_1} &=  \bra{\bf 0}\bra{\phi_2}  \left( \ket{\bf 0} A \ket{\phi_1} + \ket{\rm Garbage}\right) \\
     &= \bra{\bf 0}\bra{\phi_2} \ket{\bf 0} A \ket{\phi_1} +  \bra{\bf 0}\bra{\phi_2}\ket{\rm Garbage} \\
     &= \bra{\bf 0}\bra{\phi_2} \ket{\bf 0} A \ket{\phi_1} 
\end{split}
\end{align}
By substituting $ ket{\phi_1} = \frac{1}{||\phi_1||} \ket{\phi_1}, \ket{\phi_2} = \frac{1}{||\phi_2||} \ket{\phi_2}$ to the above, we obtain the Lemma \ref{innerproduct}. The application of Hadamard/SWAP test primitive is straightforward, provided unitary $U_A$ and state preparation routine for $\ket{\phi_1}, \ket{\phi_2}$. 

\vspace{2mm}
\noindent
\textbf{Estimating $\beta_r$ over real field $\Rbb$.} We first recall that the boundary operator is generally defined as:
\begin{equation}
\partial_r [p_0, p_1, \ldots, p_r] = \sum_{i=0}^r (-1)^i [p_0, p_1, \ldots, \hat{p_i}, \ldots, p_r],
\end{equation}
The classical knowledge of matrix $\mathscr{D}_r$ translates to the classical knowledge of $\partial_r$ directly, almost by definition. According to the above definition of boundary operator, $\partial_r$ is of size $|\Scal_{r-1}|\times |\Scal_r|$. In the $j$-th column of $\partial_r$, which corresponds to the $j$-th $r$-simplex $\sigma_{r_j}$, the location of nonzero entries are those $(r-1)$-simplexes that are faces of $\sigma_{r_j} $. The value of these non-zero entries is either 1 or -1. Thus, from $\mathscr{D}_r$, we simply need to change the value of nonzero entries at certain locations from 1 to -1. 

To proceed, we have seen earlier that the problem of estimating $r$-th Betti number, or normalized Betti number eventually reduce to estimating the kernel of the combinatorial Laplacian:
 \begin{align}
\Delta_r = \partial_{r+1} \partial_{r+1}^\dagger + \partial_r^\dagger \partial_r,
\end{align}
This has already been applied in many previous works, e.g., \cite{lloyd2016quantum, nghiem2023quantum, berry2024analyzing, schmidhuber2022complexity, ubaru2021quantum, lee2025new}. Thus, we refer the interested readers to each of them for concrete details. Here, we point out that all these works have different input assumptions and that our input model is similar to \cite{lee2025new} (to some degrees, similar to \cite{nghiem2023quantum} as well). Hence, it is most appropriate to use their result, as they also deal with a real field. We recapitulate the main procedure as well as the main result in \cite{lee2025new} with regard to the estimation of the $r$-th Betti number over the real field $\Rbb$ in the following algorithm.
\begin{method}[\cite{lee2025new}]
\label{algo: bettinumberoverrealfield}
   Let $\{ \mathscr{D}_r\}$ be the classical specification of the simplicial complex $K$ of interest, having total $N$ data points.
\end{method}
\begin{enumerate}
    \item Use the classical knowledge of $\mathscr{D}_r, \mathscr{D}_{r+1}$ and Lemma \ref{lemma: entrycomputablematrix} to construct the block-encoding of 
    $$\frac{1}{ |\Scal_{r-1}|  |\Scal_r|}  \partial_r^\dagger \partial_r, \frac{1}{ |\Scal_{r}^K|  |\Scal_{r+1}|} \partial_{r+1} \partial_{r+1}^\dagger. $$
    \item Then we use Lemma \ref{lemma: scale} to transform the block-encoded operators:
\begin{align*}
    \frac{1}{|\Scal_{r-1}|  |\Scal_r|}  \partial_r^\dagger \partial_r &\longrightarrow  \frac{ 1 }{ |\Scal_{r-1}|  |\Scal_r||\Scal_{r+1}|}\partial_r^\dagger \partial_r    \\
     \frac{1}{ |\Scal_r||\Scal_{r+1}|} \partial_{r+1} \partial_{r+1}^\dagger  &\longrightarrow   \frac{1}{ |\Scal_{r-1}|  |\Scal_r||\Scal_{r+1}|} \partial_{r+1} \partial_{r+1}^\dagger 
\end{align*}
    %The Frobenius norm of $\partial_r$ is $\sqrt{(r+1) |\Scal_r |}$. 
    \item Use the block-encoding arithmetic recipes Lemma.~\ref{lemma: sumencoding} to construct the block-encoding of their summation, resulting in the block-encoding, denoted as $U_r$, of:
    $$ \frac{1}{ 2|\Scal_{r-1}|  |\Scal_r||\Scal_{r+1}| }  \Delta_r \equiv \Delta_r' $$
        where we define a new operator $\Delta_r'$ for simplicity. Of obvious fact, the rank of $\Delta_r'$ is equal to the rank of $\Delta_r$. 
    \item Use stochastic rank estimation method proposed in \cite{ubaru2016fast, ubaru2017fast,ubaru2021quantum} (see, e.g., Algo \ref{algo: rankestimatinginifinite} in Appendix \ref{sec: randomizedrankestimation}) to estimate $\frac{\rm rank \Delta_r}{| \Scal_r |}$. Specifically:
    \begin{itemize}
    \item Draw $s$ random states from Hadamard matrix (of corresponding dimension $\dim \Delta_r$) $\ket{u_1}, \ket{u_2},..., \ket{u_s} $ with i.i.d. entries and zero mean. According to \cite{ubaru2021quantum}, this can be done by randomly applying NOT gate on $\log \dim (\Delta_r)$ qubits, followed by an application of Hadamard gates to all qubits. 
    \item Use Lemma \ref{lemma: qsvt} with $U_r$ (block-encoding of $\Delta_r'$) to obtain the block-encoding of $ T_j(\Delta_r')$ for some $j \in \Zbb_+$ where $T_j(.)$ is the $j$-th Chebyshev polynomial. 
    \item For each $i=1,2,...,s$, use Lemma \ref{innerproduct} to estimate $\braket{u_i, T_j(\Delta_r' ) u_i} $.
    \item Compute:
    \begin{align}
       \frac{1}{s}   \sum_{i=1}^s  \bra{u_i}  T_j(\Delta_r') \ket{u_i} \\
        \frac{1}{s}  \sum_{j=0}^m \sum_{i=1}^s c_j    \bra{u_i}  T_j(\Delta_r') \ket{u_i}
    \end{align}
    where $\{c_j\}_{j=1}^m$ being known factors. 
    
  \end{itemize}
\noindent
    \textbf{Output:} the estimator $ \hat{r} =  \sum_{j=0}^m \sum_{i=1}^s c_j   \bra{u_i}  T_j(\Delta_r') \ket{u_i}  $. 
    
   \textbf{Guarantee:}  Assuming that the eigenvalues of $\Delta_r'$ are $\in (\frac{1}{ \kappa_{\Delta_r'}},1)$. For $s =\mathcal{O}\left(  \frac{1}{\epsilon^2} \log \frac{1}{\delta} \right) $, and $m = \mathcal{O}\left( \log \frac{\kappa_{\Delta_r'}}{\epsilon} \right) $, then according to \cite{ubaru2016fast,ubaru2017fast}, with a failure probability $\delta$, it holds that $\big| \hat{r} - \frac{\rm rank (\Delta_r')}{\dim (\Delta_r')}\big| \leq \epsilon$. As the rank of $\Delta_r'$ is the same as the rank of $\Delta_r$, and $\dim (\Delta_r') = \dim (\Delta_r) = |\Scal_r|$ then $\hat{r}$ also satisfies:
    \begin{align}
        \big| \hat{r} - \frac{\rm rank (\Delta_r)}{|\Scal_r|}\big| \leq \epsilon
    \end{align}
   % \noindent
    %\textbf{Output:} Estimation of $r$-th normalized Betti numbers $\frac{\beta_r}{|\Scal_r|}$.
\end{enumerate}
A more completed analysis of the algorithm above can be found in \cite{lee2025new}. We recapitulate the main result in the following:
\begin{lemma}[\cite{lee2025new}]
Assuming the eigenvalues of the combinatorial $\Delta_r$ having magnitude $\in (\frac{1}{ \kappa_{\Delta_r}},1)$, which implies that the eigenvalues of 
$$ \Delta_r' \equiv \frac{1}{2|\Scal_{r-1}|  |\Scal_r||\Scal_{r+1}|}  \Delta_r $$ 
having magnitudes within
$$ (\frac{1}{ 2|\Scal_{r-1}|  |\Scal_r||\Scal_{r+1}| \kappa_{\Delta_r}},1)$$
Then the Algorithm \ref{algo: bettinumberoverrealfield} outputs an estimation of the $r$-th normalized Betti number $\frac{\beta_r}{ |\Scal_r|}$ to additive precision $\epsilon$, with success probability $1-\delta$, with time complexity
\begin{align}
\begin{split}
 &\mathcal{O}\left(\log( \abs{\Scal_{r-1}} \abs{\Scal_r}) \log\big(\kappa_{\Delta_r} |\Scal_{r-1}|  |\Scal_r||\Scal_{r+1}|\big) \frac{1}{\epsilon^2} \log \big( \frac{1}{\delta}\big) \right) \\
    &\in \mathcal{O}\left(\log( \abs{\Scal_{r-1}}\abs{\Scal_r}) \log\big(\kappa_{\Delta_r} r |\Scal_{r-1}|  |\Scal_r||\Scal_{r+1}| \big) \frac{1}{\epsilon^2} \log \big( \frac{1}{\delta}\big) \right)     
\end{split}
\end{align}
\end{lemma}
In addition, we point out that in the same work \cite{lee2025new}, the authors also proposed an alternative way to estimate Betti numbers, based on the tracking of different homology classes. \\

\noindent
\textbf{Estimating $\beta_r$ over $\mathbb{F}_p$.} We remind the readers that the boundary operator is:
\begin{equation}
\partial_r [p_0, p_1, \ldots, p_r] = \sum_{i=0}^r (-1)^i [p_0, p_1, \ldots, \hat{p_i}, \ldots, p_r],
\end{equation}
Over real field $\Rbb$, complex field $\mathbb{C}$, or integer ring $\Zbb$, $-1$ is the inverse element of $1$ in the usual sense that we use. Over finite field, however, things are different. The inverse element, say of $1$, is no longer $-1$, because in this case, for example, $\mathbb{F}_p \equiv \Zbb/p\Zbb$, that $-1$ does not even belong to this set. Instead, we need to rely on the definition provided in the Appendix \ref{sec: reviewalgebra}. It states that the inverse of an element, say $x$, of an Abelian group $\mathcal{G}$ (with group operation $+$) is another element $y$ such that $x+y =0$ where $0$ is the identity element. We consider the case $\Zbb/2\Zbb$, which contains of two element $\{0,1\}$. Then it can be seen that $1+1 = 0 $ (module 2), thus suggesting that in this group $\Zbb/2\Zbb$, $1$ is its own inverse element. More generally, let $\mathbb{F}_p \equiv (0,1,2,...,p-1)$ (with the group operation taken modulo $p$), then the inverse element of $1$ in $\mathbb{F}_p$ is $p-1$. Thus, over a finite field $\mathbb{F}_p$, the entries with value $1$ remains the same, and value $-1$ in the boundary map $\partial_r$ is replaced by $p-1$, and all the subsequent arithmetic operation needs to be done modulo $p$. However, we point out that equivalently, we can define:
\begin{align}
    \mathbb{F}_p \equiv \left( \frac{-(p-1)}{2} , \cdots, -1, 0, 1 , \cdots \frac{p-1}{2}\right) 
\end{align}
In this setup, the value of $1,-1$ remain the same, only that the arithmetic operation needs to be done modulo $p$. Therefore, it is not hard to see that the matrix representation of boundary operator remains in this finite field case.

%It makes things a bit more convenient for the representation of a boundary operator, as we will see shortly. For completeness, we use $\partial_r^{\rm mod \ p }$ to denote the boundary operator $\partial_r$ modulo $p$. 

%The first two steps of Algorithm \ref{algo: bettinumberoverrealfield} can still be applied to obtain the block-encoding of 
%\begin{align}
%    \frac{1}{ 2 \max \{|| \partial_r^{\rm mod \ p }||_F^2, \  || \partial_{r+1}^{\rm mod \ p %}||_F^2\}}  \Delta_r^{\rm mod \ p}
%\end{align}
%Previously, we have pointed out that over $\Rbb$, the Frobenius norm of $\partial_r $ is $\sqrt{(r+1)| \Scal_r|}$. Over $\mathbb{F}_p$ as defined above, as the value of $1,-1$ remain the same, then Frobenius norm is the same $|| \partial_r^{\rm mod \ p }||_F = \sqrt{(r+1)| \Scal_r|}  $ (mod p). 

The next goal is to estimate the dimension of $r$-th homology group $H_r$ over $\mathbb{F}_p$, which is also the main challenge when we change from real field $\Rbb$ to $\mathbb{F}_p$. To be more specific, at the third step of Algorithm \ref{algo: bettinumberoverrealfield}, there is an application of stochastic rank estimation method originally derived from \cite{ubaru2016fast,ubaru2017fast}, and this method only works on real, or complex field $\mathbb{C}$. Likewise, other quantum algorithms for estimating Betti numbers, such as \cite{lloyd2016quantum, berry2024analyzing, schmidhuber2022complexity}, are based on quantum simulation and quantum phase estimation, which are built naturally on complex field $\mathbb{C}$. 

To this end, we point out that there is a relevant attempt in developing classically randomized algorithms to estimate the rank of a matrix \cite{kaltofen1991wiedemann, eberly2017black} over finite field. The idea and mechanism of these algorithms are very similar to the stochastic rank estimation method proposed in \cite{ubaru2016fast,ubaru2017fast}, as they also generate a sequence of vectors with random entries. The rank of the matrix of interest, say $A$, is deducted from the action of $A$ on these randomized vectors. The only difference is that, eventually, the arithmetic operations need to be done modulo p. We leave the details of these algorithms in the Appendix \ref{sec: randomizedrankestimation}, where we will provide a concrete procedure of two algorithms to estimate rank, in the infinite case (like $\Rbb,\mathbb{C}$; see Algorithm \ref{algo: rankestimatinginifinite}) and the finite case (like $\mathbb{F}_p$; see Algorithm \ref{algo: rankestimatingfinitefield}). In the following, we formally describe our quantum algorithm for estimating the rank of a block-encoded matrix in a finite field, based on the classical algorithm Algo~\ref{algo: rankestimatingfinitefield}. 

\begin{method}
    [Quantum algorithm for estimating rank in finite field ]
    \label{algo: qaestimatingrankfinitefield}
    Let $\{ \mathscr{D}_r\}$ be the classical specification of the simplicial complex $K$ of interest, and $\mathbb{F}_p$ be the finite field of interest.
\end{method}
\begin{enumerate}
    \item Use the classical knowledge of $\mathscr{D}_r, \mathscr{D}_{r+1}$ and Lemma \ref{lemma: entrycomputablematrix} to construct the block-encoding of 
    $$ \frac{1}{ |\Scal_{r-1}|  |\Scal_r|}  \partial_r^\dagger \partial_r, \frac{1}{ |\Scal_{r}|  |\Scal_{r+1}|} \partial_{r+1} \partial_{r+1}^\dagger. $$
    Denote the above unitaries as $U_r, U_{r+1}$ respectively.
   % \item Use the block-encoding arithmetic recipes Lemma.~\ref{lemma: sumencoding} to construct the unitary block-encoding, denoted as $U_r$, of 
   % $$ \frac{ |\Scal_{r-1}|  |\Scal_r|}{ |\Scal_r|(|\Scal_{r-1}|+  |\Scal_{r+1}|)}   \frac{1}{ |\Scal_{r-1}|  |\Scal_r|}  \partial_r^\dagger \partial_r +  \frac{|\Scal_{r}|  |\Scal_{r+1}|}{ |\Scal_r|(|\Scal_{r-1}|+  |\Scal_{r+1}|)} \frac{1}{ |\Scal_{r}|  |\Scal_{r+1}|} \partial_{r+1} \partial_{r+1}^\dagger   $$
   % which
   % $$\frac{1}{ |\Scal_r|(|\Scal_{r-1}|+  |\Scal_{r+1}|)}  \Delta_r \equiv \Delta_r'$$ 

\item Use Lemma \ref{lemma: finitefieldstatepreparation} to generate $t$ different quantum states $\ket{v_1},\ket{v_2},..., \ket{v_t} $ with randomly i.i.d. entries drawn from $\mathbb{F}_p$. Likewise, generating $s$ different quantum states $ \ket{u_1},\ket{u_2},...,\ket{u_s}$ with randomly i.i.d. entries drawn from $\mathbb{F}_p$.
\item For all $i=1,2,...,t$, and $j= 1,2,...,s$, use Lemma \ref{innerproduct} to estimate the overlaps:
\begin{align}
\begin{split}
    \bra{\bf 0}  \bra{u_i} U_r \ket{\bf 0}\ket{v_j} &= \frac{1}{||u_i|| \ ||v_j||} u_i^\dagger \partial_r^\dagger \partial_r v_j \\
    &= \frac{1}{2 ||u_i|| \ ||v_j||  |\Scal_r| |\Scal_{r-1}| }  u_i^\dagger \Delta_r v_j
\end{split}
\end{align}
with a chosen additive precision 
$$\epsilon = \frac{1}{4 ||u_i|| \ ||v_j|||\Scal_r||\Scal_{r-1}| }. $$
%$$\frac{\epsilon| u_i^\dagger \Delta_r v_j | }{||u_i|| \cdot ||v_j|| 2 |\Scal_{r-1}|  |\Scal_r||\Scal_{r+1}| }. $$
Similarly, estimate the overlaps $ \bra{\bf 0} \bra{u_i} U_{r+1} \ket{\bf 0}\ket{v_j}$ to the same precision $\epsilon$. 
\item For all $i=1,2,...,t$, and $j= 1,2,...,s$, infer the value:
\begin{align}
    u_i^\dagger \partial_r^\dagger \partial_r v_j = ||u_i|| \cdot ||v_j||  2|\Scal_r|(|\Scal_{r-1}| +|\Scal_{r+1}| ) \bra{\bf 0}  \bra{u_i} U_r \ket{v_j} 
\end{align}  
By choosing 
$$\epsilon = \frac{\eta | u_i^\dagger \partial_r^\dagger \partial_r v_j  |}{4 ||u_i|| \ ||v_j|| |\Scal_r||\Scal_{r-1}| }$$
the estimation of $ u_i^\dagger \partial_r^\dagger \partial_r v_j   $ has an multiplicative accuracy $\eta$. Similarly, the estimation of $ u_i^\dagger \partial_{r+1} \partial_{r+1}^\dagger v_j$ can be estimated to the same accuracy. Then we add them to form the estimation of $ u_i^\dagger \partial_r^\dagger \partial_r v_j + u_i^\dagger \partial_{r+1} \partial_{r+1}^\dagger v_j = u_i^\dagger \Delta_r v_j $ to the same multiplicative accuracy $\eta$. 

\item Taking the estimated value above modulo $p$. Then we obtain $u_i^\dagger\Delta_r v_j  $ (mod $p$).
\item Classically build the matrix $M$ with $M_{ij} = u_i^\dagger\Delta_r v_j $ (mod $p$). Then find its rank $\rm rank (M)$ via classical Gaussian elimination modulo $p$. %For convenience, we denote $\hat{r}_p$ as the estimator of the rank of $\Delta_r^{\rm mod \ p} $. 

\noindent
\textbf{Output:} $\hat{r} = \rm rank (M)$. 

\noindent
\textbf{Guarantee: } For $s,t = \mathcal{O}\left( \log_{p} \frac{1}{\delta} \right)$ and the entries of $V,U$ are random with i.i.d. uniform over $\mathbb{F}_p$, the above estimation of $\hat{r}$ succeeds with probability:
    \begin{align}
        \rm Prob [ \hat{r} = r ] \geq  1-\delta
    \end{align}
\end{enumerate}

The quantum algorithm for torsion detection is a direct application of Algorithm \ref{algo: bettinumberoverrealfield} and \ref{algo: qaestimatingrankfinitefield}.

\begin{itemize}
    \item Repeat the Algorithm \ref{algo: qaestimatingrankfinitefield} for various values of prime $p$, we obtain a sequence of estimators $\hat{r}_{p_1}, \hat{r}_{p_2}, ...., \hat{r}_{p_K} $. 
\item Compare the values $ \hat{r}_{p_1}, \hat{r}_{p_2}, ...., \hat{r}_{p_K} $ and also the value of $\frac{\rm rank \Delta_r}{|\Scal_r|}$ estimated via Algorithm \ref{algo: bettinumberoverrealfield}. 

\noindent
\textbf{Output:} If there are differences $\longrightarrow $ complex $K$ has torsion (at order $r$). Otherwise, it does not have torsion.
\end{itemize}

\subsection{Extension to oracle-based input model}
We recall that the input information to our algorithm is the classical knowledge of matrices $\{\mathscr{D}_r\}$ encoding the mutual relation between $r$-simplexes and $(r-1)$-simplexes. This classical knowledge allows us to leverage the result of \cite{nghiem2025refined} (see Lemma \ref{lemma: entrycomputablematrix}) to block-encode the operator that is $\varpropto \Delta_r$, as well as subsequently, $\Delta_r^{\rm mod \ p}$. As mentioned earlier, existing works on quantum TDA have different input assumptions. Many, if not most of them, such as \cite{lloyd2016quantum, berry2024analyzing, hayakawa2022quantum, ubaru2021quantum}, encode the simplexes of the given complex into the computational basis states of appropriate qubits system. For instance, assuming that the complex $K$ has $N$ data points. Let $v_1,v_2,...,v_N$ denotes these $N$ data points, or 0-simplexes. Then a 1-simplex $[v_i,v_j]$ is encoded to the computational basis state of $N$-qubits system, that has the bit at position $i$ and $j$ equal to $1$, and 0 otherwise. Likewise, a 2-simplex $[v_i,v_j,v_k]$ is encoded to the basis state that has the bit $1$ at position $i,j,k$, and 0 otherwise. Generally, a $r$-simplexes is encoded into a string of Hamming weight $r+1$. In addition, another enabling component of these algorithms is the oracle $O_r$, which acts on a string $\sigma_r$ of Hamming weight $r+1$ as follows:
\begin{align}
    O_r \ket{0} \ket{\sigma_r} = \begin{cases}
        \ket{1} \ket{\sigma_r} \text{\ if $\sigma_r \in $ K  }\\
        \ket{0} \ket{\sigma_r} \text{ otherwise} 
    \end{cases}
\end{align}
One can see that the above oracle is similar to the oracle in the Grover algorithm. In fact, the algorithm in \cite{lloyd2016quantum, berry2024analyzing, hayakawa2022quantum, ubaru2021quantum} employs a multi-solution version of Grover's search algorithm as a subroutine. 

The integration of the above oracle with our algorithm is quite straightforward as a corollary of a technique introduced \cite{hayakawa2022quantum}. Specifically, in Section 6, they show the following:
\begin{lemma}[Section 6 of \cite{hayakawa2022quantum}]
    Provided the oracle $O_r$ as above, then there is a quantum circuit of complexity $\mathcal{O}(N^2)$ that is an exact block-encoding of $ \frac{\partial_r}{N(r+1)}$. 
\end{lemma}
Thus, we can use the above result to obtain the block-encoding of:
\begin{align}
     \frac{\partial_r}{N(r+1)},  \frac{\partial_r^\dagger}{N(r+1)},  \frac{\partial_{r+1}}{N(r+2)},   \frac{\partial_{r+1}^\dagger}{N(r+2)}
\end{align}
%The above result allows us to estimate $ u_i^\dagger \Delta_r v_j$ more conveniently. We recall $\Delta_r = \partial_r^\dagger %\partial_r + \partial_{r+1}\partial_{r+1}^\dagger$, and thus:
%\begin{align}
%    u_i^\dagger \Delta_r v_j &= u_i^\dagger (  \partial_r^\dagger \partial_r + \partial_{r+1}\partial_{r+1}^\dagger ) v_j \\
%    &= u_i^\dagger \partial_r^\dagger \partial_r v_j    + u_i^\dagger\partial_{r+1}\partial_{r+1}^\dagger v_j 
%\end{align}
%Provided the block-encoding of $ \frac{\partial_r}{N(r+1)},  \frac{\partial_{r+1}}{N(r+2)}$, we can use Lemma \ref{innerproduct} to estimate: 
%\begin{align}
%    \frac{1}{N^2(r+1)^2} u_i^\dagger \partial_r v_j
%\end{align}
Then we use Lemma \ref{lemma: product} to construct the block-encoding of
$$ \frac{1}{N^2(r+1)^2} \partial_r^\dagger \partial_r , \frac{1}{N^2 (r+2)^2} \partial_{r+1} \partial_{r+1}^\dagger  $$
%\begin{align}
%    \frac{1}{2N^2 \left( (r+1)^2 +(r+2)^2 \right) }\left( \partial_r^\dagger \partial_r + \partial_{r+1} \partial_{r+1}^\dagger %\right) =  \frac{1}{2N^2 \left( (r+1)^2 +(r+2)^2 \right)} \Delta_r
%\end{align}
From here, can execute a similar procedure as Algo.~\ref{algo: qaestimatingrankfinitefield} to find out the rank of $\Delta_r$ over various $p$, then compare them in order to detect torsion.

\section{Complexity and Error Analysis}
\label{sec: complexityanalysis}
In this section, we provide a detailed discussion on the overall error and complexity of the algorithm outlined above (the non-oracle case). We will follow exactly each step of the algorithm \ref{algo: qaestimatingrankfinitefield} and explicitly detail the complexity of the recipes used in each step. An analysis for the oracle caes will be given afterward.  \\

\subsection{Analysis for non-oracle case}

\noindent
\textbf{Step 1:} We first point out that the operator $\mathscr{D}_r$ has size $|\Scal_{r-1}| \times |\Scal_r| $. Thus, an application of Lemma \ref{lemma: entrycomputablematrix} uses a quantum circuit of depth, respectively
\begin{align}
    \mathcal{O}\left( \log  |\Scal_{r-1}||\Scal_r|  \right) ,\mathcal{O}\left( \log  |\Scal_{r+1}||\Scal_r|  \right) 
\end{align}

%\noindent
%\textbf{Step 2:} Next, we use Lemma \ref{lemma: sumencoding} to combine the block-encoding of two operators, each with circuit %depth as above. So the total circuit depth is:
%\begin{align}
%    \mathcal{O}\left( \log  |\Scal_{r-1}||\Scal_r| |\Scal_{r+1}|  )\right) 
%\end{align}

\noindent
\textbf{Step 2:} At this step, we use Lemma \ref{lemma: finitefieldstatepreparation} to prepare the states (of dimension $|\Scal_r|$) with randomly i.i.d. entries drawn from $\mathbb{F}_p$, which use a quantum circuit of depth:
\begin{align}
    \mathcal{O}\left( \log |\Scal_r| \right) 
\end{align}

\noindent
\textbf{Step 3 \& 4 \& 5:} In the first two steps, we need to use the block-encoding from the previous two steps, inside the Hadamard/SWAP test primitive, to estimate the overlaps:
\begin{align}
\begin{split}
    \bra{\bf 0}  \bra{u_i} U_r \ket{\bf 0}\ket{v_j} 
    &= \frac{1}{||u_i|| \ ||v_j|| |\Scal_r|( |\Scal_{r-1}|  +|\Scal_{r+1}|)}   u_i^\dagger \partial_r^\dagger \partial_r v_j
\end{split}
\end{align}
with an additive precision 
$$  \epsilon = \frac{\eta |u_i^\dagger  \partial_r^\dagger \partial_r  v_j  |}{4 ||u_i|| \ ||v_j|| |\Scal_r| |\Scal_{r-1}| }$$
%$$\frac{ \delta | u_i^\dagger \Delta_r v_j | }{ ||u_i|| \cdot ||v_j|| 2 |\Scal_{r-1}|  |\Scal_r||\Scal_{r+1}| }. $$
which implies an estimation of $  u_i^\dagger  \partial_r^\dagger \partial_r v_j$ with multiplicative accuracy $\eta$. Because $ u_i^\dagger  \partial_r^\dagger \partial_r v_j $ is an integer, it can be inferred from the estimation above by choosing $\eta = \frac{1}{2}$. For example, with such a multiplicative precision, we simply need to round the estimation above to the nearest integer. We note that the primitives above use amplitude estimation as a subroutine. As such, we can use the recent results \cite{rall2021faster,rall2023amplitude,aaronson2020quantum} which bypass the need of the quantum Fourier transform compared to the traditional methods. Provided that the circuit depth of the block-encoding of 
$$  \frac{   \partial_r^\dagger \partial_r }{|\Scal_r| |\Scal_{r-1}|}$$ 
and of $\ket{u_i}, \ket{v_j} $ are, respectively:
\begin{align}
        \mathcal{O}\left( \log | \Scal_{r-1}| |\Scal_r|    \right) , \ \mathcal{O}\left(\log |\Scal_r|  \right) 
\end{align}
%Thus, circuit complexity of estimating $\bra{\bf 0}  \bra{u_i} U_r \ket{\bf 0}\ket{v_j} $, to an additive precision $\epsilon$ is:
%\begin{align}
%      \mathcal{O}\left( \frac{1}{\epsilon}  \log | \Scal_{r-1}| |\Scal_r|   \right) 
%\end{align}
%Then we infer the value of $u_i^\dagger \Delta_r v_j $ as 
%\begin{align}
%    u_i^\dagger \Delta_r v_j = 2 ||u_i|| \ ||v_j|| |\Scal_{r-1}|  |\Scal_r||\Scal_{r+1}| \bra{\bf 0} % \bra{u_i} U_r \ket{\bf 0}\ket{v_j} 
%\end{align}
The estimation above has circuit complexity
$$ \mathcal{O}\left( \log | \Scal_{r-1}| |\Scal_r|  \frac{1}{\eta} \sqrt{\frac{2 ||u_i|| \ ||v_j|| |\Scal_r| |\Scal_{r-1}|   }{ u_i^\dagger  \partial_r^\dagger \partial_r v_j }}    \right).   $$
We remark on the following property:
\begin{align}
    \frac{ u_i^\dagger  \partial_r^\dagger \partial_r v_j }{ ||u_i|| \ ||v_j|| } = \bra{u_i}  \partial_r^\dagger \partial_r \ket{v_j}
\end{align}
Under the promise that $\partial_r$ (and hence, $ \partial_r^\dagger \partial_r$) has eigenvalues bounded, the above quantity can be safely treated as $\mathcal{O}(1)$. Therefore, the circuit complexity above is simplified further as:
$$ \mathcal{O}\left( \log \left( | \Scal_{r-1}| |\Scal_r| \right) \frac{1}{\eta} \sqrt{ |\Scal_r|(|\Scal_{r-1}|  }   \right). $$
Similarly, the estimation of $u_i^\dagger \partial_{r+1}\partial_{r+1}^\dagger v_j$ to the same multiplicative accuracy $\eta$ has complexity
$$ \mathcal{O}\left( \log \left( | \Scal_{r+1}| |\Scal_r|\right)  \frac{1}{\eta} \sqrt{ |\Scal_r| |\Scal_{r+1}|  }   \right). $$

For simplicity, we define $S_r^{\max} = \max \{ |\Scal_{r+1}| , |\Scal_r|,  \Scal_{r-1}|  \} $. The total complexity in the above estimations is:
$$ \mathcal{O}\left( \log \left( S_r^{\max} \right)  \frac{1}{\eta}  S_r^{\max}     \right) $$

As the last \textbf{Step 5}, taking modulo $p$ of $u_i^\dagger \Delta_r v_j $. As we need to evaluate  $ u_i^\dagger\Delta_r v_j$ (mod $p$) for $i=1,2,...,t$, $j=1,2,...,s$, so we need to repeat the above process $s\cdot t = \mathcal{O}\left( ( \log_p \frac{1}{\delta})^2\right)$ times.
%, resulting in total complexity 
%\begin{align}
%      \mathcal{O}\left( \frac{1}{\epsilon} ( \log_p \frac{1}{\delta})^2 \log | \Scal_{r-1}| |\Scal_r|  %\frac{||u_i|| \cdot ||v_j|| 2 |\Scal_{r-1}|  |\Scal_r||\Scal_{r+1}| }{ \epsilon | u_i^\dagger %\Delta_r v_j |}     \right) 
%\end{align}

\noindent
\textbf{Step 6\& 7 \& 8:} Performing Gaussian elimination on a matrix of size $m \times n$ has complexity $\mathcal{O}(mn)$ \cite{dumas2012computational}. So the complexity for finding the rank of $M$ (mod $p$) is $\mathcal{O}\left( st \right) \in \mathcal{O} \left(( \log_p \frac{1}{\delta})^2 \right) $. 

\vspace{{2mm}}
\noindent
\textbf{Summary.} To sum up, for a fixed multiplicative accuracy $\eta$, our algorithm makes use of a quantum circuit of maximum complexity
\begin{align}
      \mathcal{O}\left( \log \left( S_r^{\max} \right)  \frac{1}{\eta}   S_r^{\max}     \right) 
\end{align}
and need to repeat such circuit $\mathcal{O}\left( \log^2_p \frac{1}{\delta} \right) $ times, assisted by a classical routine of complexity 
$$\mathcal{O}\left( \log^2_p \frac{1}{\delta} \right). $$

\subsection{Analysis for oracle case}
\noindent
While the above complexity analysis is for the non-oracle case, the complexity analysis for the oracle case can be carried out in a straightforward manner. The only difference is that, in the \textbf{Step 1}, we use the result of \cite{hayakawa2022quantum} to obtain the block-encoding of $ \frac{\partial_r}{N (r+1)}$. The complexity of these two steps is $\mathcal{O}\left( N^2 \right)$. From the \textbf{Step 2} onward, the procedure is the same, with merely the interchange between $ |\Scal_r| |\Scal_{r-1}| \leftrightarrow N^2(r+1)^2$ and $ |\Scal_r| |\Scal_{r+1}| \leftrightarrow N^2(r+2)^2$. Thus the complexity can be deduced in a straightforward manner. To sum up, we have that the quantum circuit complexity in the oracle case is:
\begin{align}
    \mathcal{O}\left( \big(  N^2 + \log |\Scal_r| \big)  \frac{1}{\eta } Nr \right) 
\end{align}
%\begin{align}
%    \mathcal{O}\left( N^2  \kappa_{\Delta_r}^2  \log^4 \left( \frac{N(r+1)(r+2)}{\epsilon} \right) %\right) = \mathcal{O}\left( N^2\kappa_{\Delta_r} ^2 \log^4 \frac{Nr}{\epsilon}  \right)
%\end{align}
%The remaining steps and corresponding complexity can be deduced in a straightforward manner.  The %final circuit depth is: 
%\begin{align}
%    \mathcal{O}\left(  \left( N^2\kappa_{\Delta_r}^2  \log^4 \big(Nr  \big) + \log |\Scal_r|\right) %\kappa_{\Delta_r } \max_{i,j}\left(||u_i|| \cdot ||v_j|| \right) \right)
%\end{align}
The total of iterations is $st = \mathcal{O}\left(  \big( \log_p \frac{1}{\delta} \big)^2\right)$ for a success probability $1-\delta$, followed by a classical algorithm of complexity
$$\mathcal{O}\left(  \log^2_p \frac{1}{\delta} \right). $$

\section{Classical Randomized Algorithm for Rank Estimation}
\label{sec: randomizedrankestimation}
For convenience, we provide a neat summary, or pipeline of the two algorithms for rank estimation over infinite and finite field, respectively. We refer the interested readers to the correspondingly original works for a more rigorous treatment. 

The first one is the stochastic rank estimation proposed in \cite{ubaru2016fast,ubaru2017fast,ubaru2021quantum}. The method is based on the property that $  \rm rank A \equiv \Tr \left( h(A) \right)$ where $h(x)$ is a step function for $x \in (-1,1)$. The step function is then approximated by Chebyshev polynomial $h(x) \approx \sum_{j=0}^m c_j T_j(x)$ where $T_j(.)$ is the Chebyshev polynomial. 
\begin{method}[Rank estimation over $\Rbb/\mathbb{C}$ \cite{ubaru2016fast,ubaru2017fast}]
\label{algo: rankestimatinginifinite}
    Let $A \in \Rbb^{n \times n}/\mathbb{C}^{n \times n}$ be a Hermitian matrix with eigenvalues having magnitude $\in (\frac{1}{\kappa_A},1)$. 
\end{method}
\begin{enumerate}
    \item Draw $s$ random vectors $u_1,u_2,...,u_s \in \Rbb^n$ with i.i.d. entries $\in \Rbb/ \mathbb{C}$. 
    \item For each $i=1,2,...,s$, compute $\braket{u_i, T_j(A) u_i} $ where $T_j(.)$ is the Chebyshev polynomial. 
    \item Compute:
    \begin{align}
       \frac{1}{s}   \sum_{i=1}^s  u_i^\dagger  T_j(A) u_i \\
        \frac{1}{s}  \sum_{j=0}^m \sum_{i=1}^s c_j   u_i^\dagger  T_j(A) u_i 
    \end{align}
    \noindent
    \textbf{Output:} $ \hat{r} =  \sum_{j=0}^m \sum_{i=1}^s c_j  u_i^\dagger  T_j(A) u_i   $ is the estimator for the rank of $A$. 

    \noindent
    \textbf{Guarantee:} For $s =\mathcal{O}\left(  \frac{1}{\epsilon^2} \log \frac{1}{\delta} \right), m = \mathcal{O}\left( \log \frac{\kappa_A}{\epsilon} \right) $, with a success probability $\geq 1-\delta$, it holds that:
    \begin{align}
        \big| \hat{r} - r \big| \leq \epsilon
    \end{align}
    \noindent
    \textbf{Time complexity:}  $\mathcal{O}\left( n^2 \frac{1}{\epsilon^2} \log \big( \frac{1}{\delta }\big) \log \frac{\kappa_A}{\epsilon}   \right) $
\end{enumerate}
As can be seen, the above method is based on the approximation of $\rm rank(A)$ as a trace of certain function of $A$ (the step function). The trace is then approximated by a polynomial of corresponding matrix. An important ingredient in the above algorithm is the first step. As pointed out in \cite{ubaru2016fast,ubaru2017fast}, any random vector with zero mean and i.i.d. coordinates can work. To obtain such a sequence of random vectors, the authors in \cite{ubaru2021quantum} proposed to randomly draw columns from the Hadamard matrix. As emphasized in \cite{ubaru2021quantum}, this choice works very well in practical setting. To achieve this, they began with the state $\ket{0}^{\log n}$, followed by a random application of NOT gate in each of the $\log n$ qubits, and then Hadamard gates to all $\log n$ qubits. Repeating this process $s$ times results in random Hadamard states $\ket{u_1}, \ket{u_2}, ..., \ket{u_s}$, which is sufficient for the purpose of trace and rank estimation in the above algorithm. To be more precise, since the states $\ket{u_1},\ket{u_2},...,\ket{u_s}$ are normalized, the estimation $\frac{1}{s}  \sum_{j=0}^m \sum_{i=1}^s c_j   \bra{u_i}T_j(A) \ket{u_i}  $ actually approximate $\frac{\Tr A}{N}$.

It can be seen that the above discussion does not hold in the finite case, as we cannot approximate the trace and rank via Chebyshev polynomials. To this end, it turns out that there are multiple classical algorithms tailored for finite case. In the following, we provide a summary/pipeline of the method introduced in \cite{kaltofen1991wiedemann, eberly2017black}.
\begin{method}[Rank Estimation Over $ \mathbb{F}_p$ \cite{kaltofen1991wiedemann, eberly2017black} ]
\label{algo: rankestimatingfinitefield}
    Let $A \in \mathbb{F}_p^{n \times n}$ be a matrix of size $m \times n$ over the field $\mathbb{F}_p$. Let $r$ denotes the rank of $A$.  
\end{method}
\begin{enumerate}
    \item Draw $t$ random vectors $v_1,v_2,...,v_t \in \mathbb{F}_p^n$. Define $V$ be a matrix of size $n \times t$, with $[v_1,v_2,...,v_t] $ as columns.
    \item Draw $s$ random vectors $u_1,u_2,...,u_s \in \mathbb{F}_p^n$. Define $U$ be a matrix of size $s \times n$, with $ [u_1^T,u_2^T,...,u_s^T]$ as rows.
    \item Define a matrix $M$ of size $s \times t$ with entries:
    \begin{align}
        M_{ij} = u_i^T A v_j  \text{ \ (mod p)}
    \end{align}
    Or equivalently
    \begin{align}
        M = U A V
    \end{align}
    \item Find rank $\hat{r}$ of $M$ (via Gaussian elimination mod p). 
    
    \noindent
    \textbf{Output:} $\hat{r}$ is an estimator for the rank of $M$. 
    
    \noindent
    \textbf{Guarantee: } For $s,t = \mathcal{O}\left( \log_{p} \frac{1}{\delta} \right)$ and the entries of $V,U$ are random with i.i.d. uniform over $\mathbb{F}_p$, the above estimation of $\hat{r}$ succeeds with probability:
    \begin{align}
        \rm Prob [ \hat{r} = r ] \geq  1-\delta
    \end{align}
    \textbf{Time complexity:}  $\mathcal{O}\left( n^2 \log \frac{1}{\delta }   \right) $
\end{enumerate}

\section{Proof of Lemma \ref{lemma: entrycomputablematrix} and Lemma \ref{lemma: finitefieldstatepreparation}}
\label{sec: proofoflemmaentrycomputablematrix}
We remark that Lemma \ref{lemma: entrycomputablematrix} is the result of \cite{lee2025new}, which is based on \cite{nghiem2025refined}. The original proof can thus be found in Appendix C of \cite{lee2025new}. Here, for completeness, we directly quote their proof to aid the interested readers. 

\subsection{General framework \cite{nghiem2025refined}}
As mentioned, Lemma \ref{lemma: entrycomputablematrix} can be seen as a corollary of a recent result \cite{nghiem2025refined}, which generally shows that it is possible to block-encode a matrix (up to a scaling factor) provided its classical description, or its entries. First, we point out the following property. Let $A$ be a matrix of size $M\times N$. Let $A^i$ denotes the $i$-th column of $A$. Then consider the following vector (not necessarily normalized)
\begin{align}
     \frac{1}{\sqrt{MN}}\sum_{i=1}^N \underbrace{A^i}_{\rm register 1}\ket{i}
\end{align}
is of dimension $MN$. If we trace out the first register in the above vector, we obtain the following operator:
\begin{align}
  \frac{1}{MN} A^\dagger A
\end{align}
Our goal is to construct the block-encoding of the above operator, provided the classical knowledge, or the entries of $A$. Before describing the procedure, we mention a helpful lemma:
\begin{lemma}[\cite{gilyen2019quantum} Block Encoding Density Matrix]
\label{lemma: improveddme}
Let $\rho = \Tr_A \ket{\Phi}\bra{\Phi}$, where $\rho \in \mathbb{H}_B$, $\ket{\Phi} \in  \mathbb{H}_A \otimes \mathbb{H}_B$. Given unitary $U$ that generates $\ket{\Phi}$ from $\ket{\bf 0}_A \otimes \ket{\bf 0}_B$, then there exists a highly efficient procedure that constructs an exact unitary block encoding of $\rho$ using $U$ and $U^\dagger$ a single time, respectively.
\end{lemma}
%Provided the classical knowledge of $A$, which means that the entries of $A^i$ are known for all $i$, the state $\ket{A}$ can be prepared using any of the state preparation protocol \cite{grover2000synthesis,grover2002creating,plesch2011quantum, schuld2018supervised, nakaji2022approximate,marin2023quantum,zoufal2019quantum,prakash2014quantum, zhang2022quantum}. 

%The complexity of the block encoding of $ \frac{1}{||A||_F^2} A^\dagger A $ depends mainly on the complexity of the unitary that prepares $\ket{A}$, as the subsequent step, e.g., Lemma \ref{lemma: improveddme}, would make use of this unitary once more. The method in \cite{zhang2022quantum} is universal, which means that it can be applied to any kind of states and arguably achieve the optimal depth. If $\ket{A}$ is prepared via this method, then the complexity, including the quantum circuit depth required is $\mathcal{O}(\log sN)$ , the number of ancilla qubits $\mathcal{O}(sN)$ and classical pre-processing time $\mathcal{O}(\log MN)$. In practice, this approach is only qubit-efficient when the sparsity $s$ is not large. On the other hand, if $\ket{A}$ has any of the structure as in \cite{grover2000synthesis,grover2002creating,plesch2011quantum, schuld2018supervised, nakaji2022approximate,marin2023quantum,zoufal2019quantum,prakash2014quantum}, then it can be prepared, with circuit complexity, including total number of gates, qubits, and classical pre-processing time varied case by case. 

\noindent
\textbf{Approach 1 -- based on \cite{mcardle2022quantum}.} The state preparation method in \cite{mcardle2022quantum} achieves efficient depth while using a modest number of ancilla at the same time. Specifically, according to the description below the Theorem 1 in \cite{mcardle2022quantum}, let $f:[-a,a] \longrightarrow \Rbb$ be a (preferably smooth) function of arbitrary parity, and define:  
$$\ket{\Phi_f} = \frac{1}{\mathcal{N}_f}\sum_{ i=-\frac{N}{2}}^{\frac{N}{2 } -1} f \left( \frac{2a i}{N} \right)  \ket{i} $$
where $\mathcal{N}_f = \sqrt{ \sum_{i} f(.)^2}$. Then there exists a deterministic procedure (see Figure 1 of \cite{mcardle2022quantum}) that prepares the state:
\begin{align}
    \frac{1}{\sqrt{N}}\ket{00}\sum_{i}^{N} f \left( \frac{2a i}{N}\right) \ket{i}+ \ket{\rm Garbage}  
\end{align}
where $\ket{\rm Garbage}$ refers to the redundant part that is completely orthogonal to the first part $ \ket{00}\sum_{i}^{N} f \left( \frac{2a i}{N}\right) \ket{i}$. The quantum circuit complexity of the above procedure is $\mathcal{O}\left( \rm deg(f) \log N \right) $ (where $\deg(f)$ is the degree of function $f$), with extra 3 ancilla qubits. We remark that in the above, it was mentioned that the function $f$ being smooth. As discussed in \cite{mcardle2022quantum}, their method can also be simply extended to non-smooth functions as well, with asymptotically the same complexity. To obtain the state $\ket{\Phi_f}$, we measure the ancilla and post-select $\ket{00}$. The success probability of this measurement is $\mathcal{O}\left( \frac{\mathcal{N}_f  }{N} \right)$. By using oblivious amplitude amplification, which incurs the quantum circuit complexity by a factor $ \sim \sqrt{\frac{N}{\mathcal{N}_f}}$, the success probability of the measurement is close to unity. Therefore, the total circuit complexity for obtaining $\ket{\Phi_f}$ is $\mathcal{O}\left(\sqrt{\frac{N}{\mathcal{N}_f}}  \rm deg(f) \log N\right) $.

To apply the above result, we first promote the new variable $k = j+(i-1)N$, which has the property: $\ket{k }= \ket{j+(i-1)N} = \ket{j}\ket{i}$. Then there is a correspondence between states $ \sum_{i,j=1}^{N,M} A_{ij} \ket{j} \ket{i} = \sum_{k=1}^{N^2} a_k \ket{k}$ where the new amplitude $a_k \equiv a_{j +(i-1)N} = A_{ji}$. For any (preferably smooth) function $f: [-1,1] \longrightarrow [-1,1]$, if the entries $\{ a_k \}_{k=1}^{MN}$ obey $a_k = f(\frac{k}{MN}) = f \big(j + (i-1)N\big) = A_{ji} $, then as mentioned earlier, there exists a deterministic procedure that prepares the state:
\begin{align}
\begin{split}
     &\frac{1}{\sqrt{MN}} \ket{00}\sum_{k=1}^{MN} f( \frac{k}{MN}) \ket{k}+ \ket{\rm Garbage} \\
    &= \frac{1}{\sqrt{MN}} \ket{00}\sum_{k=1}^{MN} a_k\ket{k}+ \ket{\rm Garbage} \\
   % &= \frac{1}{\sqrt{MN}} \ket{00}\sum_{k=1}^{MN} f(k)\ket{k}+ \ket{\rm Garbage} \\
    &= \frac{1}{\sqrt{MN}}\ket{00}\sum_{i,j=1}^{N,M} A_{ji} \ket{j} \ket{i}+ \ket{\rm Garbage}  \\
    &=   \frac{1}{\sqrt{MN}}\ket{00} \sum_{i}^N A^i\ket{i} + \ket{\rm Garbage}
\end{split}
    \label{c3}
\end{align}
The above state can be achieved using a quantum circuit having gate complexity $\mathcal{O}\left( \rm deg(f) \log MN \right) $ (where $\deg(f)$ is the degree of function $f$) and 3 ancilla qubits. We comment that in the above, we have assumed that there is a function $f$ satisfying the stated condition. In principle, there can be many possible choices of $f$. Subsequently, in Appendix \ref{sec: proofofstatepreparation} (see the paragraph above Appendix \ref{sec: analysisofthenorm}), we will discuss an effective route to obtain the function $f$ as desired, provided the classical knowledge of $\{A_{ij}\}$.  

To obtain the block-ending of $\sim A^\dagger A$ from the above state, we append another ancilla initialized in $\ket{00}$ to the state above, so we have the state:
\begin{align}
    \frac{1}{\sqrt{MN}} \sum_{i}^N \underbrace{\ket{00}}_{\rm register 1}\underbrace{\ket{00}}_{\rm register 2} A^i\ket{i} + \ket{00}\ket{\rm Garbage}
\end{align}
followed by two CNOT gates between register 1 and 2. Then we obtain the state: 
\begin{align}
    \frac{1}{\sqrt{MN}} \ket{00} \sum_{i}^N \underbrace{\ket{00} A^i}_{\rm register X }\ket{i} +  \ket{\rm Garbage'}\ket{\rm Garbage}
\end{align}
where $\ket{\rm Garbage'} $ is completely orthogonal to $\ket{00}$. 
As we pointed out earlier, if we trace out the first register of the vector $ \frac{1}{\sqrt{MN}} \sum_{i}^N  A^i \ket{i}$, then we obtain the operator $\frac{1}{MN} A^\dagger A $. As such, if we trace out the \rm{register X}  in the state above, we obtain the density state:
\begin{align}
    \ket{00}\bra{00} \otimes \frac{1}{MN}  A^\dagger A + (...)
\end{align}
where (...) refers to the redundant part. The density state above can be block-encoded via the Lemma \ref{lemma: improveddme}. According to Definition \ref{def: blockencode}, this density state is also the block-encoding of $ \frac{1}{MN}  A^\dagger A $, thus implying that we have achieved the desired block-encoding. 

Recall that the quantum circuit complexity of preparing the state in Eqn.~\ref{c3} is $\mathcal{O}\left( \log MN  \right) $ (we assume $\deg(f) = \mathcal{O}(1)$). The adding of two more ancilla qubits $\ket{00}$ and usage of 2 CNOT gates increases the circuit complexity by $\mathcal{O}(1)$. Finally, the application of Lemma \ref{lemma: improveddme} uses the state preparation unitary in Eqn.~\ref{c3} one more, thus the total circuit complexity is $\mathcal{O}\left( \log MN\right)$. 

\vspace{2mm}
\noindent
\textbf{Approach 2 -- based on \cite{marin2023quantum, grover2002creating}.} In the above procedure, we used the state preparation method in \cite{mcardle2022quantum} to prepare the state that contains $ \sim \sum_{i,j=1}^{N,M} A_{ji} \ket{j}\ket{i}$ as a subvector. However, we remark that other state preparation methods exist. For example, the method in \cite{grover2002creating, marin2023quantum} considers the state of the form $ \sum_{i=1}^N f(i) \ket{i}$ where $f:[0,1]: \longrightarrow \mathbb{R}_+$ is a continuous, positively integrable function satisfying $  \sum_{i=1}^N f(i)^2 = 1$. They show that if $f(.)$ is efficiently integrable, then $ \sum_{i=1}^N f(i) \ket{i}$ can be prepared, with a fidelity $\epsilon$, using a $\log N$-qubit quantum circuit of complexity $\mathcal{O}\left( 2^{k(\epsilon)}\right) $, where 
$$ k(\epsilon) = \max  \{  -\frac{1}{2} \log_2 \big( 4^{-\log N} - 96  \log (1-\epsilon) \big) , 2 \}  $$ 
which is asymptotically independent of $N$, according to \cite{marin2023quantum}. 

To apply this method, first we define $||A||_F = \sqrt{\sum_{i,j}A_{ij}^2}$ and define the new variable $k = j + (i-1)N$ as above. Suppose that the entries $\{A_{ij}\}_{i,j=1}^N$ are positive (we will show how to deal with negative entries subsequently), and there is some positive integrable function $f:[0,1] \longrightarrow (0,1]$ such that: $f(\frac{k}{MN}) = f(\frac{j + (i-1)N}{MN}) = \frac{1}{||A||_F} A_{ji}$, then the method of \cite{marin2023quantum} can be used to prepare the  state 
\begin{align}
    \sum_{k=1}^{MN} f(\frac{k}{MN}) \ket{k}  = \sum_{i,j=1}^{M,N} \frac{1}{||A||_F} A_{ji} \ket{k}   =  \frac{1}{||A||_F} \sum_{i,j=1}^{N,M} A_{ji} \ket{j}\ket{i}.
\end{align}
%$$ \sum_{k=1}^{MN} f(\frac{k}{MN}) \ket{k}  = \sum_{i,j=1}^{M,N} \frac{1}{||A||_F} A_{ji} \ket{k}   =  \frac{1}{||A||_F} \sum_{i,j=1}^{N,M} A_{ji} \ket{j}\ket{i}.$$
Again, we recall what we commented earlier, that in reality there can be many possible choice for $f$. One way to obtain a function $f$ as desired will be discussed in Appendix \ref{sec: proofofstatepreparation} (see paragraph above Appendix \ref{sec: analysisofthenorm}).

Given the procedure that prepares the state above, then by executing the same procedure as below Eqn.~\ref{c3}, the block-encoding of $\frac{1}{||A||_F^2} A^\dagger A$ can be obtained. As the last and optional step, an application of Lemma \ref{lemma: scale} with the scaling factor $ \frac{||A||_F^2}{MN}$ can be used to transform the block-encoded operator 
$$\frac{1}{||A||_F^2} A^\dagger A \longrightarrow \frac{1}{MN} A^\dagger A. $$
Provided that the circuit complexity for preparing $\frac{1}{||A||_F} \sum_{i,j=1}^{N,M} A_{ij} \ket{j}\ket{i} $ is $\mathcal{O}\left(  MN\right)$, and Lemma \ref{lemma: improveddme} uses this state preparation procedure 1 more time, the complexity for obtaining the block-encoding of the operator above is $\mathcal{O}\left( \log MN \right) $. 

%We point out an important subtlety. As mentioned above, the state preparation procedure in \cite{marin2023quantum} requires the function $f$ to be positive. It implies that it can only be applied to prepare $ \frac{1}{||A||_F} \sum_{i,j=1}^{N,M} A_{ij} \ket{j}\ket{i}$ if $A_{ij} \geq 0$ for all $i,j$. Subsequently, we will show how to extend this method to deal with the case where $A$ contains negative entries. 

Now we turn our attention to the case where $A$ contains both positive and negative entries. 
Define $||A^+||_F = \sqrt{\sum_{ij, A_{ij} \geq 0} A_{ij}^+ }$ and $|||A^-||_F =\sqrt{\sum_{ij, A_{ij} < 0} A_{ij}^2} $. We consider state: 
\begin{align}
\begin{split}
     \ket{\phi_+ } &= \frac{1 }{ ||A^+||_F} \sum_{ij, A_{ij} \geq 0} A_{ij} \ket{j}\ket{i} \\
     \ket{\phi_-} &=  \frac{1 }{ ||A^-||_F} \sum_{ij, A_{ij} < 0} A_{ij} \ket{j}\ket{i} 
\end{split}
\end{align}
The state $\ket{\phi_+}$ has nonnegative entries, so it can be efficiently prepared via the procedure mentioned earlier. Let $U_+$ denotes the unitary that prepare this state $\ket{\phi_+}$, i.e.,, $U_+ \ket{0}^{ \log MN} = \ket{\phi_+}$. Regarding the state $\ket{\phi_-}$, we consider the state $ - \ket{\phi_-} = \frac{1 }{ ||A^-||_F} \sum_{ij, A_{ij} < 0} (-A_{ij}) \ket{j}\ket{i}  $. Because $ A_{ij} < 0$, so $-A_{ij} >0$, which is positive. Therefore, the state $ -\ket{\phi_-}$ can be prepared, with the corresponding preparation unitary $U_-$. Then it can be seen that the first column of $U_+$ and $U_-$ are $ \ket{\phi_+}$ and $-\ket{\phi_-}$, respectively. Lemma \ref{lemma: sumencoding} can be used to construct the block-encoding of 
\begin{align}
    \frac{||A^+||_F }{||A||_F } U_+ -  \frac{||A^-||_F }{||A||_F }U_- 
\end{align}
which has the first column to be: 
\begin{align}
   \frac{||A^+||_F }{||A||_F }  \ket{\phi_+} + \frac{||A^-||_F }{||A||_F }\ket{\phi_-}  = \frac{1}{||A||_F} \left(\sum_{ij, A_{ij} \geq 0} A_{ij} \ket{j}\ket{i} +  \sum_{ij, A_{ij} < 0} A_{ij} \ket{j}\ket{i}\right). 
\end{align}
which is exactly $  \frac{1}{||A||_F} \sum_{i,j=1}^{N,M} A_{ij} \ket{j}\ket{i}$. By applying this unitary to $\ket{0}^{\log MN}$, we obtain the desired state. Then we can proceed the same procedure as outlined earlier to obtain the block-encoding of $\frac{1}{MN} A^\dagger A $, with the complexity being the same, of order $\mathcal{O}\left( \log MN\right)$.

In addition, we remark that in the same work \cite{marin2023quantum}, the authors also proposed a variational approach to prepare the desired state. As empirically demonstrated, their variational approach can work well in practice without a lot of training time.

\vspace{2mm}
\noindent
\textbf{Approach 3 -- based on \cite{nakaji2022approximate}.}  At the same time, the method in \cite{nakaji2022approximate} proposes another variational approach to prepare a state of the form $ \frac{1}{||x||}\sum_{i=1}^N x_i \ket{i}$ where $||x||^2 = \sum_{i=1}^N x_i^2$. As analyzed and numerically verified in \cite{nakaji2022approximate}, a parameterized quantum circuit of complexity $\mathcal{O}\left( l \log N\right) $ (where $l = \mathcal{O}(1)$) is sufficient to prepare the desired state, even without error tolerance. In addition, the maximum number of ancilla qubits is $\mathcal{O}(1)$. Therefore, a direct application of this state preparation method allows us to prepare the state $ \frac{1}{||A||_F} \sum_{i,j=1}^{N,M} A_{ij} \ket{j}\ket{i}$. The procedure for obtaining the block-encoding $\frac{1}{||A||_F^2}A^\dagger A$, and then $\frac{1}{MN} A^\dagger A $ is straightforward described in the previous paragraph, achieving the same complexity. 

\begin{figure}[t!]
    \centering
        \begin{tikzpicture}
            
\draw[->] (-1.4,0) -- (6,0) node[below] {$x$};
\draw[->] (-1,-1) -- (-1,5) node[left] {$f(x)$};
%\draw[blue, thick, domain=-1.3:1.8, smooth, variable=\x] plot (\x, {\x^3 - 2*\x + 2});
\filldraw[black] (-1,1) circle (2pt);
\node[left] at (-1,1) {$x_1$};
\filldraw[black] (0,3) circle (2pt);
\node[left] at (0,3) {$x_2$};
\filldraw[black] (1,2) circle (2pt) ;
\node[left]  at (1,2) {$x_3$};
\filldraw[black] (2, 2) circle (2pt);
\node[above] at (2,2) {$x_4$};
\filldraw[black] (3,0) circle (2pt);
\node[below] at (3,0) {$x_5$};
\filldraw[black] (4,-1) circle (2pt);
\node[below] at (4,-1) {$x_6$};

\draw[-] (-1,1) -- (0,3);
\draw[-] (0,3) -- (1,2) ;
\draw[-] (1,2) -- (2,2);
\draw[-] (2,2) -- (3,0);
\draw[-] (3,0) -- (4,-1);
%$\foreach \i in {-1,0,1,2, 3 }
 %   \fill (\i,{\i^3 - 2*\i + 2}) circle (2pt)  node[above right] { };
        \end{tikzpicture}

    \caption{Simple illustration of the general idea behind our state preparation for $\ket{\Phi}$ indicated at the beginning. We promote $x_i$ to a point on the $(x,y)$ plan with its $x$ coordinate being the index $i$ of $x_i$, and $y$ being the value $x_i$. Es each amplitude $x_i$ is randomly chosen from $\mathbb{F}_p$, it can be seen that $ \frac{-p+1}{2}  \leq x_i \leq \frac{p-1}{2}  $. By connecting these points $\{x_i\}_{i=1}^6$, a piecewise-linear function is formed. }
    \label{fig:mainfigure}
\end{figure}
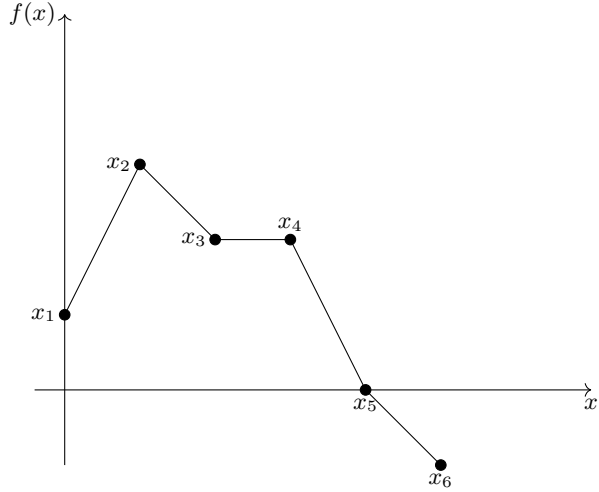

%\vspace{2mm}
%\noindent
%\textbf{Remark:} Within the \textbf{Approach 1} and \textbf{Approach 2} above, to apply the methods in \cite{mcardle2022quantum} and \cite{marin2023quantum}, respectively, we need to choose the function $f$ accordingly. In principle, any choice of $f$ satisfying the values at $\{ ij \}$ works, e.g., $f(ij) = A_{ij}$ in the first approach,  $f(ij/MN) = A_{ij}/||A||_F$ in second approach. As such, a possible choice of $f$ is the piecewise-linear function on $\mathbb{R}^2$ satisfying $f(ij) = A_{ij}$ (for \textbf{Approach 1}) and $f(\frac{ij}{MN}) = \frac{1}{||A||_F} A_{ij} $ (for \textbf{Approach 2}). See Figure \ref{fig:mainfigure} for a simple illustrative example. 

%We remark that this function, despite being piecewise-linear, is still continuous, and thus the rigorous performance guaranty of the method in \cite{marin2023quantum} still holds. Beside, as we mentioned earlier, the method in \cite{mcardle2022quantum} can also handle non-smooth function, which includes the piecewise-smooth function as in our case. 

\subsection{Application to our case: $A \equiv \partial_r$} 
Now we return to the context of Lemma \ref{lemma: entrycomputablematrix}, where the role of matrix $A$ above is replaced by the boundary operator $\partial_r$, which has dimension $\dim \partial_r = |\Scal_{r-1}| |\Scal_r|$. By defining an arbitrary (preferably smooth) function that satisfies 
$$f(\frac{j + (i-1)N}{|\Scal_{r-1}| |\Scal_r|}) = (\partial_r)_{ij} \in \{-1,0,1\}$$,
then the method of \cite{mcardle2022quantum} as well as the procedure outlined above is directly applicable so as to obtain the block-encoding of $ \frac{1}{ \dim \partial_r} \partial_r^T \partial_r$. %Because the size of $\partial_r$ is $|S_{r-1}^K| \times |S_r^K|$, its dimension $\dim \partial_r = |S_{r-1}^K| |S_r^K|$. 
This block-encoding procedure has total circuit complexity $\mathcal{O}\left( \log |S_r||S_{r-1}| \right)$ and in particular, using $\mathcal{O}(1)$ ancilla qubits only. 

On the other hand, instead of using \cite{mcardle2022quantum}, we can use the state preparation method in \cite{grover2002creating, marin2023quantum} by defining a positive integrable function that has values 
$$f(\frac{j + (i-1)N}{|\Scal_{r-1}|\ |\Scal_r|}) =  \frac{1}{||\partial_r||_F} (\partial_r)_{ij} \in \{  -\frac{1}{||\partial_r||_F} ,0,\frac{1}{||\partial_r||_F}   \}. $$
The circuit complexity of this approach is $\mathcal{O}\left( 2^{k(\epsilon)} \right) $ with $k(\epsilon)$ defined earlier, and in addition, this approach does not use any ancilla qubits. Last, the variational approach of \cite{marin2023quantum} or  \cite{nakaji2022approximate} can also be applied, using a quantum circuit of complexity $\mathcal{O}\left( \log |\Scal_r{-1}| |\Scal_r|\right)$ and maximally extra $\mathcal{O}(1)$ ancilla qubits.

Thus, overall, we have completed the proof of Lemma \ref{lemma: entrycomputablematrix}. In the following, we will proceed to the proof of Lemma \ref{lemma: finitefieldstatepreparation}. In particular, we will show a route to enhance the state preparation methods previously discussed. 

\subsection{Proof of Lemma \ref{lemma: finitefieldstatepreparation} and a followed-up byproduct}
\label{sec: proofofstatepreparation}
Recall that our goal is to construct a unitary that prepares the following state:
\begin{align}
    \ket{\Phi} = \frac{1}{||\xbf||} \sum_{ i=0}^{N-1} x_i \ket{i}
\end{align}
where $||a\| = \sqrt{\sum_{ i=0}^{N-1} a_i^2}$ and each $a_i$ is probabilistically drawn from the finite set:
\begin{align}
    \mathbb{F}_p \equiv \left( \frac{-(p-1)}{2} , \cdots, -1, 0, 1 , \cdots \frac{p-1}{2}\right) 
\end{align}
with $p$ being a prime number. It is easy to see that once the values of $\{x_i\}$ are provided, this is just the corollary of the state preparation methods discussed earlier. In the following, we discuss two important practical aspects of state preparation methods \cite{mcardle2022quantum} and \cite{marin2023quantum}. 

First, as we described earlier, the method of \cite{mcardle2022quantum} involves ancilla measurement and post-selection. Thus, the overall complexity contains the factor $\mathcal{O}\left( \frac{\sqrt{N}}{||\xbf||} \right) = \mathcal{O}\left(  \sqrt{\frac{N}{||\xbf||^2}}\right)  $, which can be large if $|| \xbf||$ is small. In the next appendix, we will provide a more detailed analysis, showing that the expectation value of $||\xbf||^2 $ is of order $\mathcal{O}(N)$ and thus guaranties the complexity of this step to be $\mathcal{O}(1)$, on average.

Second, to apply the method of \cite{mcardle2022quantum} and \cite{marin2023quantum}, we need a function that, respectively, satisfies $f(\frac{i}{N}) = x_i$ (for \cite{mcardle2022quantum}) and $f(\frac{x}{N}) = \frac{x_i}{||\xbf||}$ (for \cite{marin2023quantum}). In practice, we are provided with values of $x_i$ only. The challenge then is: how to effectively choose the function $f$ as desired ? Although there can be many choices, and hence, probably many ways to find a good function, we propose the following route. First, we promote the points $\{ x_i \}_{i=1}^N$ to the $x-y$ plane, with corresponding coordinates $\{ (i, x_i) \}_{i=1}^N$. We then connect these points consecutively from lower indexes to higher ones. The result is a piecewise-linear function, e.g., see Fig~\ref{fig:mainfigure} for a simple illustration with 6 points. Then we can take advantage of existing results in approximating the piecewise-linear function \cite{lipman2010approximating, jimenez2016transforming, arandiga2005interpolation} to obtain the function $f$ as desired. Therefore, we have provided a justification for what we have stated earlier. In the context of \textbf{Approach 1} and \textbf{Approach 2}, we can applied the route we just described to find the desired $f$, so as to prepare the state $ \sim \sum_{i,j=1}^{N,M} A_{ij}\ket{j}\ket{i}$. 

As a final remark, piecewise-linear function is still continuous, and thus the rigorous performance guaranty of \cite{marin2023quantum} still holds. In addition, piecewise-linear function in our case is also piecewise-smooth, so the method \cite{mcardle2022quantum} can still be applicable. As a byproduct, what we describe above is an effective way to expand the capability of state preparation methods \cite{marin2023quantum} and \cite{mcardle2022quantum} in practice.

\section{Analysis of the norm $ ||\xbf|| $}
\label{sec: analysisofthenorm}
We begin with the following statement, with its proof provided subsequently:
\begin{lemma}
\label{lemma: bound}
    Let $\{ x_i\}_{i=1}^N$ be i.i.d. entries randomly drawn from $\mathbb{F}_p \equiv \{ -\frac{p-1}{2},-\frac{p-1}{2}+1, ..., -1,0,1, \frac{p-1}{2}+1, \frac{p-1}{2}  \} $ with probability $\frac{1}{p}$. Define $||\xbf|| = \sqrt{\sum_{i=1}^N x_i^2}$. Then we have 
    \begin{itemize}
        \item (Weak version) For some $S \in \Rbb_+$ and $S^2 \geq \frac{N(p^2-1)}{12}$, the probability that $||\xbf|| \leq S$ is:
        \begin{align}
            \rm Pr(||\xbf|| \leq S )  \geq 1- \frac{N  \left(\frac{p^4}{180} - \frac{p^2}{36} + \frac{1}{45} \right)^2 }{N  \left(\frac{p^4}{180} - \frac{p^2}{36} + \frac{1}{45} \right)^2 + (S^2 - \frac{N(p^2-1)}{12})^2}
        \end{align} 
        \item (Stronger version) For any $S \in \Rbb_+$:
        \begin{align}
            \rm Pr(||\xbf|| \leq S )  \geq  \Phi \left(  \frac{S^2 - \frac{N(p^2-1)}{12}}{ \sqrt{N}  \left(\frac{p^4}{180} - \frac{p^2}{36} + \frac{1}{45} \right)}\right) - \frac{0.56 (p-1)^6}{2^6  \left(\frac{p^4}{180} - \frac{p^2}{36} + \frac{1}{45} \right)^3 \sqrt{N}}
        \end{align}
    \end{itemize}
    where $\Phi$ is the standard normal CDF.
\end{lemma}
\noindent
\textbf{Proof:} Our proof makes use of well-known results from probability theory $\&$ statistics. We define the following variable:
\begin{align}
    x_i &\in \mathbb{F}_p \\
    Z_i &= x_i^2 \\
    ||\xbf||^2 &= \sum_{i=1}^N Z_i
\end{align}
For convenience, we denote $\mathbb{F}_p \equiv \{ -\frac{p-1}{2} +k  \}$ for $k=0,1,...,p-1$. Then it can be seen that:
\begin{align}
    \rm Pr( x_i = -\frac{p-1}{2} +k) = \frac{1}{p}
\end{align}
for any $i =1,2,...,N$ and $k=0,1,...,p-1$. We compute the mean and the variance of $Z_i$ as follows:
\begin{align}
    \mathbb{E}( Z_i) &= \sum_{k=0}^{p-1} \rm Pr( x_i = -\frac{p-1}{2} +k) \left( -\frac{p-1}{2} +k \right)^2 \\
    &= \sum_{k=0}^{p-1}\frac{1}{p} \left( -\frac{p-1}{2} +k \right)^2 \\
    &= \frac{p^2-1}{12} \\
    \rm Var (Z_i) &= \sum_{k=0}^{p-1}  \rm Pr( x_i = -\frac{p-1}{2} +k)\frac{ \left( \left( -\frac{p-1}{2} +k \right)^2 - \mathbb{E}( Z_i) \right)^2}{N-1} \\
    &=  \frac{p^4}{180} - \frac{p^2}{36} + \frac{1}{45}
\end{align}
With regard to $||\xbf||^2 = \sum_{i=1}^N Z_i $, we have the following:
\begin{align}
    \mathbb{E}(||\xbf||^2  ) &=  \sum_{i=1}^N \mathbb{E}( Z_i)  \\
    & = N \frac{p^2-1}{12} \\
    \rm Var (||\xbf||^2 )  &= N \sum_{i=1}^N \rm Var (Z_i) \\ &= N \left( \frac{p^4}{180} - \frac{p^2}{36} + \frac{1}{45}\right)
\end{align}
We also point out that:
\begin{align}
    \rm Pr ( ||\xbf|| \leq S )  = \rm Pr ( ||\xbf||^2 \leq S^2). 
\end{align}
The weak version of the lemma above is a consequence of the one-sided Chebysev-Cantelli inequality, which states that, for a random variable $Y$ drawn from a probability distribution with mean $\mu$ and variance $\sigma$, it holds that:
\begin{align}
    \rm Pr ( Y \geq \mu + t) \leq \frac{\sigma^2}{ \sigma^2 +t^2} 
\end{align}
In our case, we need to replace:
\begin{align}
    Y &= ||\xbf||^2 \\
    \mu &=\mathbb{E}(||\xbf||^2  )  =  N \frac{p^2-1}{12}\\
    \sigma^2 &= \rm Var (||\xbf||^2 ) =N \left( \frac{p^4}{180} - \frac{p^2}{36} + \frac{1}{45}\right) \\
    t &= S^2 - \mu 
\end{align}
Then we have:
\begin{align}
    \rm Pro( ||\xbf||^2 \geq S^2 )  \leq  \frac{ \sigma^2 }{ \sigma^2 + \big( S^2- \mathbb{E}(||\xbf||\big)^2  }
\end{align}
which implies that:
\begin{align}
    \rm Pro( ||\xbf||^2  \leq S^2 )  &\geq 1- \frac{\sigma^2}{ \sigma^2 + \big( S^2- \mathbb{E}(||\xbf||\big)^2 } \\
    &\geq 1- \frac{N  \left(\frac{p^4}{180} - \frac{p^2}{36} + \frac{1}{45} \right)^2 }{N  \left(\frac{p^4}{180} - \frac{p^2}{36} + \frac{1}{45} \right)^2 + (S^2 - \frac{N(p^2-1)}{12})^2}
\end{align}
thus completing the proof of the weak version of Lemma \ref{lemma: bound}. 

The strong version, on the other hand, is a consequence of the Berry-Esseen-type normal bound. It states that for a random variable $Y$ drawn from a probability distribution with mean $\mu$, variance $\sigma^2$, defining the variable $X = \frac{S^2- \mu}{\sigma}$, then it holds that:
\begin{align}
    \rm Pr(Y \leq S^2) \geq \Phi(x) - \frac{C \rho}{\sigma^3/N}
\end{align}
where $C \leq 0.56$ is a constant, and $\rho \equiv \mathbb{E} \big( | Z_i - \mathbb{E}(Z_i) |\big)^3$. Again, we replace:
\begin{align}
    Y &= ||\xbf||^2 \\
    \mu &=\mathbb{E}(||\xbf||^2  )  =  N \frac{p^2-1}{12}\\
    \sigma^2 &= \rm Var (||\xbf||^2 ) =N \left( \frac{p^4}{180} - \frac{p^2}{36} + \frac{1}{45}\right) 
\end{align}
Now we consider specifically the term $\rho \equiv \mathbb{E} \big( | Z_i - \mathbb{E}(Z_i) |\big)^3 $, showing that it is upper bounded. We have that:
\begin{align}
    \mathbb{E} \big( | Z_i - \mathbb{E}(Z_i) |\big)^3 &= \sum_{k=0}^{p-1} \rm Pr( x_i = -\frac{p-1}{2} +k) \Big| \left(-\frac{p-1}{2} +k) \right)^2  - \frac{p^2-1}{12}  \Big|^3 \\
    &= \sum_{k=0}^{p-1} \frac{1}{p} \Big| \left(-\frac{p-1}{2} +k) \right)^2  - \frac{p^2-1}{12}  \Big|^3
\end{align}
We observe that, for all $k$:
\begin{align}
    \left(-\frac{p-1}{2} +k) \right)^2  - \frac{p^2-1}{12}  \leq \left(\frac{p-1}{2} \right)^2
\end{align}
Thus:
\begin{align}
    \sum_{k=0}^{p-1} \frac{1}{p} \Big| \left(-\frac{p-1}{2} +k) \right)^2  - \frac{p^2-1}{12}  \Big|^3 &\leq \sum_{k=0}^{p-1} \frac{1}{p} \left(\frac{p-1}{2}\right)^6 \\
    &\leq \left(\frac{p-1}{2}\right)^6
\end{align}
So we have:
\begin{align}
    \rho \equiv \mathbb{E} \big( | Z_i - \mathbb{E}(Z_i) |\big)^3 \leq \left(\frac{p-1}{2}\right)^6
\end{align}
So from the inequality:
\begin{align}
    \rm Pr(Y \leq S^2) \geq \Phi(x) - \frac{C \rho}{\sigma^3/N}
\end{align}
we can deduce that:
\begin{align}
    \rm Pr (||\xbf||^2 \leq S^2) \geq  \Phi \left(  \frac{S^2 - \frac{N(p^2-1)}{12}}{ \sqrt{N}  \left(\frac{p^4}{180} - \frac{p^2}{36} + \frac{1}{45} \right)}\right) - \frac{0.56 (p-1)^6}{2^6  \left(\frac{p^4}{180} - \frac{p^2}{36} + \frac{1}{45} \right)^3 \sqrt{N}}
\end{align}
which is inherently equivalent to:
\begin{align}
     \rm Pr (||\xbf|| \leq S) \geq  \Phi \left(  \frac{S^2 - \frac{N(p^2-1)}{12}}{ \sqrt{N}  \left(\frac{p^4}{180} - \frac{p^2}{36} + \frac{1}{45} \right)}\right) - \frac{0.56 (p-1)^6}{2^6  \left(\frac{p^4}{180} - \frac{p^2}{36} + \frac{1}{45} \right)^3 \sqrt{N}}
\end{align}
%Combining the above bounds and the complexity from previous section, we have that the probability for $\max_i \{ ||u_i||\}, \max_j \{ ||v_j ||\}$, and thus $\mathcal{C}$, having bounded $l_2$-norms obey the above inequality, with $N$ replaced by $|\Scal_r|$. 

\end{document}